\documentclass[useAMS,usenatbib,usegraphicx]{mn2e}
\usepackage{times}

\def\sun{\odot}

\title[VLBA Observations of Mrk\,6: Probing the Jet-Lobe Connection]{VLBA Observations of Mrk\,6: Probing the Jet-Lobe Connection}
\author[P. Kharb et al.]{P. Kharb,$^{1}$\thanks{E-mail:kharb@iiap.res.in} C. P. O'Dea,$^{2}$ S. A. Baum,$^3$ M. J. Hardcastle,$^4$ D. Dicken,$^{5}$  
\newauthor J. H. Croston$^6$, B. Mingo$^{4,7}$ and J. Noel-Storr$^3$ \\ 
$^{1}$Indian Institute of Astrophysics, II Block, Koramangala, Bangalore 560034, India\\
$^{2}$School of Physics and Astronomy, Rochester Institute of Technology, Rochester, NY 14623\\
$^3$Chester F. Carlson Center for Imaging Science, Rochester Institute of Technology, Rochester, NY 14623\\
$^4$School of Physics, Astronomy and Mathematics, University of Hertfordshire, College Lane, Hatfield AL10 9AB, UK\\
$^5$CEA-Saclay, F-91191 Gif-sur-Yvette, France\\
$^6$School of Physics and Astronomy, University of Southampton, Southampton, Hampshire SO17 1BJ, UK\\
$^7$Department of Physics and Astronomy, University of Leicester, University Road, Leicester LE1 7RH, UK}

\begin{document}
\date{Accepted . Received ; in original form }
\pagerange{\pageref{firstpage}--\pageref{lastpage}} \pubyear{2014}
\maketitle
\label{firstpage}

\begin{abstract}
We present the results of high resolution VLBI observations at 1.6 and 4.9~GHz of the radio-loud Seyfert galaxy, Mrk\,6. These observations are able to detect a compact radio core in this galaxy for the first time. The core has an inverted spectral index ($\alpha^{1.6}_{4.9}$=+1.0$\pm$0.2) and a brightness temperature of $1\times10^8$\,K. Three distinct radio components which resemble jet elements and/or hot spots, are also detected. The position angles of these elongated jet elements point, not only to a curved jet in Mrk\,6, but also towards a connection between the AGN and the kpc-scale radio lobes/bubbles in this galaxy. Firmer constraints on the star formation rate {provided by} new {\it Herschel} observations (SFR $<0.8$~M$_{\sun}$~yr$^{-1}$) make the starburst-wind powered bubble scenario implausible. From plasma speeds obtained via prior {\it Chandra} X-ray observations, and ram pressure balance arguments for the ISM and radio bubbles, the north-south bubbles are expected to take {$7.5\times10^6$~yr} to form, and the east-west bubbles $1.4\times10^6$ yr. We suggest that the jet axis has changed at least once in Mrk\,6 within the last {$\approx10^7$~yr}. A comparison of the nuclear radio-loudness of Mrk\,6 and a small sample of Seyfert galaxies with a subset of low-luminosity FRI radio galaxies reveals a continuum in radio properties. 
\end{abstract}

\section{Introduction}
{Markarian\,6 is a nearby\footnote{$z$ = 0.018813, luminosity distance = 80.7 Mpc} bulge-dominated lenticular galaxy hosting a Seyfert 1.5 type active galactic nucleus (AGN)}. Lenticular galaxies are considered to be intermediate between spiral and elliptical galaxies; they have a disk but have either consumed or lost most of their interstellar medium (ISM), so that there is little ongoing star formation activity. {At radio frequencies,} Mrk\,6 exhibits a highly complex morphology: two pairs of self-similar bipolar bubble-like radio structures {that are} oriented nearly perpendicular to each other {are clearly visible} \citep{Kharb06}. While starburst superwinds have been suggested to be {mainly responsible for} the creation of lobe/bubble-like structures in Seyfert galaxies \citep[e.g.,][]{Baum93}, several pieces of evidence point towards an AGN {origin} \citep[e.g.,][]{Colbert96,Elmouttie98,Gallimore06,HotaSaikia06,Kharb06,Kharb10a}. 

The complex radio morphology of Mrk\,6, along with a high degree of polarization ($\sim50$\% at the edges) and an AGN jet-like spectrum {($\alpha\sim-0.8$; defined throughout as $S_\nu\propto \nu^{\alpha}$),} led \citet{Kharb06} to suggest that the edge-brightened bubbles were the projections of precessing radio jets. The precessing jets had to shut-off and restart in a nearly perpendicular direction to account for both sets of bubbles. The dominant role played by the AGN in the creation of the radio bubbles was consistent with the low star-formation rate (SFR) of $\le5.5$~M$_\sun$~yr$^{-1}$ derived from the {\it Spitzer} Space Telescope data of \citet{Buchanan06}: an SFR of $\approx33$~M$_\sun$~yr$^{-1}$ was required for the bubbles to have been inflated by supernova explosions. New {\it Herschel} telescope data place even more stringent constraints on the SFR, as discussed ahead.

High-resolution radio observations with the MERLIN\footnote{Multi-Element Radio-Linked Interferometer Network} interferometer were able to identify individual components in the north-south kpc radio jet \citep{Kukula96}. As the MERLIN observations were carried out at a single frequency, \citet{Kukula96} could not identify a flat spectrum radio core in Mrk\,6. On the basis of the similar position angles (PAs) of the jet and the brightest regions along the edges of the large north-south lobes, they suggested that these bright regions were hot spots created by the north-south jet, where most of the bulk energy of the jets was dissipated; the jet plasma then drifted along the local host galaxy density gradient along the minor axis, forming the rest of the radio lobes. This of course cannot explain the east-west bubbles in this source, because they lie close to the major axis of the galactic bulge {(and the galactic disk)}, where the density gradient is not so pronounced. However, since Kukula et al. did not actually detect the full extent of the east-west bubbles in their observations, this question did not arise. The precessing jet scenario invoked by \citet{Kharb06} suggested on the other hand, that the jet did not dissipate all its energy in these hot spots but rather continued to advance beyond them to create the 8-shaped radio structures. Furthermore, the precession model indicated that the north-south kpc jet was more closely associated with the east-west bubbles, rather than the north-south bubbles \citep[see Figure~6,][]{Kharb06}. 

In order to re-examine these issues, as well as to search for a flat or inverted spectrum radio core, we observed Mrk\,6 with the Very Large Baseline Array (VLBA) at 1.6 and 4.9~GHz. We present here the results from these observations. We also discuss results from two epochs of archival 8.4~GHz VLBA data on Mrk\,6. We have used radio bubbles and lobes interchangeably in the text. However, the lobes in Seyfert galaxies differ from those found in powerful {Fanaroff-Riley type II radio galaxies \citep[FRII;][]{Fanaroff74}} in that they are edge-brightened on all sides \citep[e.g.,][]{Mingo12} and not just at the leading edges like in FRIIs \citep[e.g.,][]{Kharb08}. 

At the distance of Mrk\,6, $1\arcsec$ corresponds to 377 pc, for $H_0$ = 71 km~s$^{-1}$~Mpc$^{-1}$, $\Omega_{mat}$ = 0.27, $\Omega_{vac}$ = 0.73. 

\section{Observations and Data Reduction}
We observed Mrk\,6 at 1.65 and 4.98 GHz in a phase-referencing experiment \citep{BeasleyConway95} on 20th October, 2006 (Project ID BK132), with the ten elements of the VLBA \citep{Napier94} for a total time of $\approx$12 hours. The total on-source time was $\approx$240 minutes at 1.6 GHz and $\approx$310 minutes at 4.9 GHz. The calibrator DA193 was used as the fringe-finder, while the compact source J0639+7324 with a switching angle of 1.33$\degr$ and an X, Y positional error of 0.16, 0.22 {milliarcsec (mas)}, respectively, was chosen as the phase reference calibrator\footnote{http://www.vlba.nrao.edu/astro/calib/}. A switching cycle of $\approx$12 mins (calibrator 2 mins, target 10 mins) was adopted at both the frequencies. The calibrator J0621+7605 ($\sim2.65\degr$ away and with an X, Y positional uncertainty of 5.81, 2.80 mas) was used as the phase check source. The phase check source is used to get a measure of the decorrelation expected on the target in case of a non-detection.

The observations\footnote{http://www.vlba.nrao.edu/astro/VOBS/astronomy/oct06/bk132/} were made using the standard setup with 128 Mbps recording, a one bit sampling rate, and two 8 MHz intermediate frequencies (IFs) in dual polarization. This yielded a total bandwidth of ($8\times2\times2$=) 32 MHz. We followed the standard VLBA data reduction procedure using the NRAO Astronomical Image Processing System \citep[AIPS,][]{Greisen03} as described in Appendix C of the AIPS Cookbook. Delays were corrected using fringe fitting on the fringe finder (with AIPS task FRING). The phase calibrator J0639+7324 was iteratively imaged and self-calibrated on both phase and phase+amplitude using AIPS tasks IMAGR and CALIB. The images were then used as models to determine the amplitude and phase gains for the antennas. These gains were applied to the target and the final images were made using IMAGR. 

The standard self-calibration procedure was not possible for the weak radio components in Mrk\,6 at either frequency: the task CALIB had too many failed solutions, irrespective of the weighting and/or uvranges selected. Therefore, phase errors could not be consistently corrected, resulting in the scattering and a net reduction of source flux density. The images displayed in Figures~\ref{figwhole} and \ref{figcband} were created with uniform weighting (with ROBUST=5 in IMAGR). The final rms noise in the image was $\sim$1.5$\times10^{-4}$~Jy~beam$^{-1}$ at 1.6~GHz and $\sim$6.7$\times10^{-5}$~Jy~beam$^{-1}$ at 4.9~GHz. 

In order to obtain the 1.6$-$4.9~GHz spectral indices, the uv-coverage at both the frequencies was constrained to be identical. The 1.6 and 4.9~GHz images were created using a UVRANGE of 3 - 48\,M$\lambda$ in task IMAGR. Additionally, a UVTAPER at 32\,M$\lambda$ was applied to the 4.9~GHz image. The 1.6~GHz image was created using a ROBUST parameter of $-0.3$. This matching of the shortest and longest baselines resulted in 1.6 and 4.9~GHz images having synthesized beam-sizes close to 5~mas. Therefore, prior to estimating the spectral {indices} for individual jet components, the 1.6 and 4.9~GHz images were convolved with circular beams of size 5~mas. The spectral indices of components S1 and S2 were finally obtained from integrated flux density values derived from identical-sized regions using the task JMFIT, following the relation, $\alpha$ = log$(S_{\nu_1}/S_{\nu_2})$/log$(\nu_1/\nu_2)$.    

We also reduced two epochs of {archival} 8.36 GHz VLBA data on Mrk\,6 from 14th and 24th March, 2011 (Project IDs BC196K, BC196L; {PI Jim Condon}). {J0639+7324} was the phase reference calibrator {for Mrk\,6}. The total {on-source time} was only $\approx$5 mins. After calibrating in the standard manner and imaging, we failed to detect any emission from Mrk\,6 at either epoch. The final rms noise in the images, {which was close to the expected thermal noise for this experimental setup, was} $\sim2.0\times10^{-4}$ Jy~beam$^{-1}$, and $\sim3.1\times10^{-4}$ Jy~beam$^{-1}$, for the March 14 and March 24 data, respectively\footnote{http://www.vlba.nrao.edu/astro/VOBS/astronomy/apr11/bc196m/}.

{The new {\it Herschel} data discussed ahead in Section~4, were obtained from the {\it Herschel} Science Archive from the program $\mathrm{OT1}\_\mathrm{rmusholtz}\_1$. Both the PACS and SPIRE data were processed with the {\it Herschel} HIPE analysis software (version 10.0.0). The PACS data were acquired in March 2012 in the standard photometry mode  at wavelength filters 70 and 100$\mu$m and were reduced with default scripts for extended sources. The SPIRE data were taken in September 2011 {at 250, 350 and 500$\mu$m} in the small scan mode and were reduced in HIPE using scripts developed for optimum processing at the Institut d$'$Astophysique Spatiale. Fluxes were measured using the HIPE aperture photometry tool and the uncertainty reported is the calibration uncertainty combined in quadrature with the variation in the background measured, at random, with identical apertures.}

\section{Results}
\begin{figure}
\centering{\includegraphics[width=15cm]{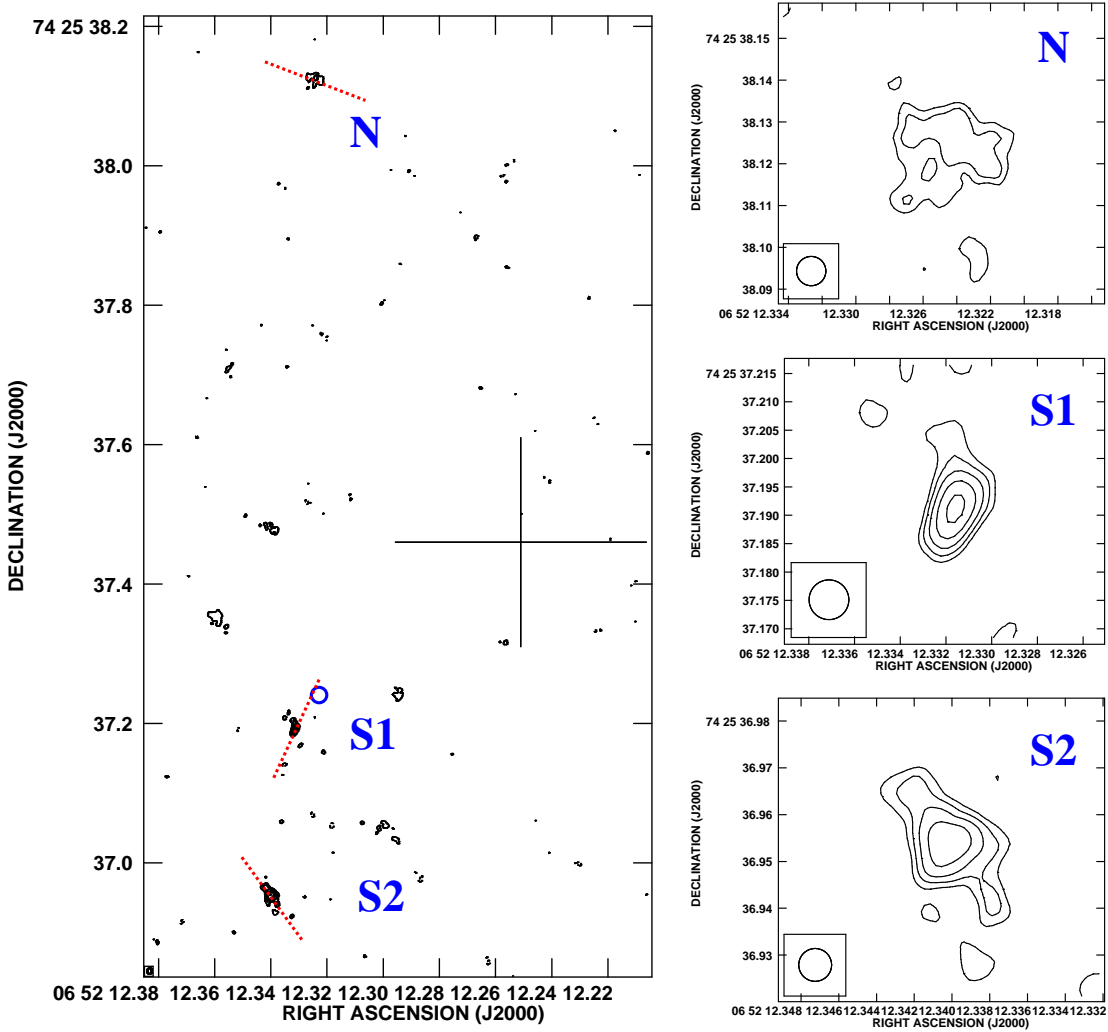}}
\caption{The 1.6~GHz radio image of Mrk\,6: three components N, S1, and S2 are detected. The panels on the right show the zoomed-in images of the components. The red dashed lines indicate the PA of the component extensions. The contours for the right hand side panels are in percentage of peak intensity and increase in steps of $\sqrt{2}$. The peak intensity is 2.2 mJy~beam$^{-1}$, and the lowest contour levels are for (N) $\pm$16\%, (S1) $\pm$22.5\%, and (S2) $\pm$22.5\%. The images were convolved with circular beams of size 7~mas\,$\times$\,7~mas. The cross represents the centre of the optical host galaxy of Mrk\,6 along with the uncertainty, while the small blue circle represents the position of the radio core seen only at 4.9~GHz.}
\label{figwhole}
\end{figure}

\begin{figure}
\centering{
\includegraphics[width=10cm]{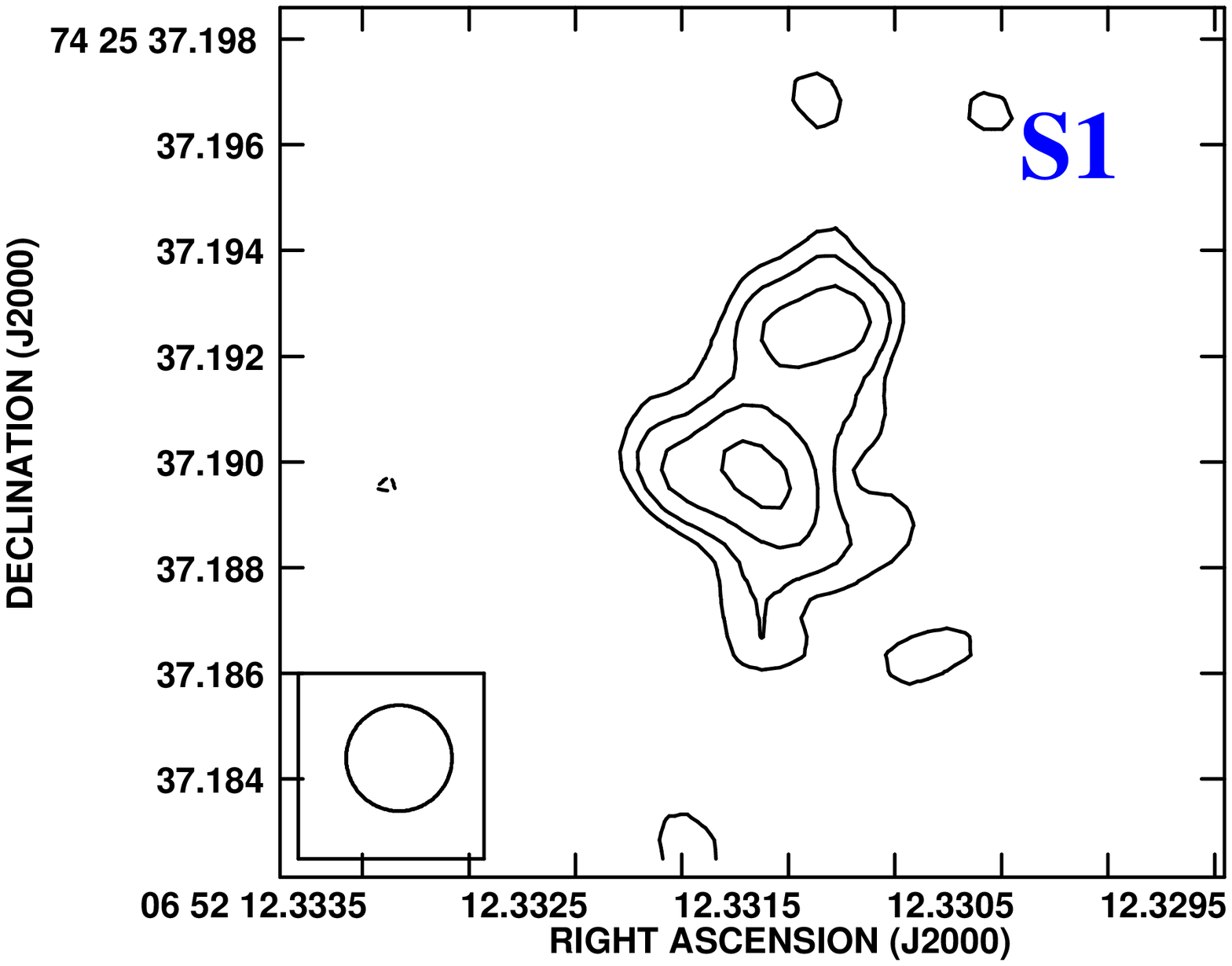}
\includegraphics[width=10cm]{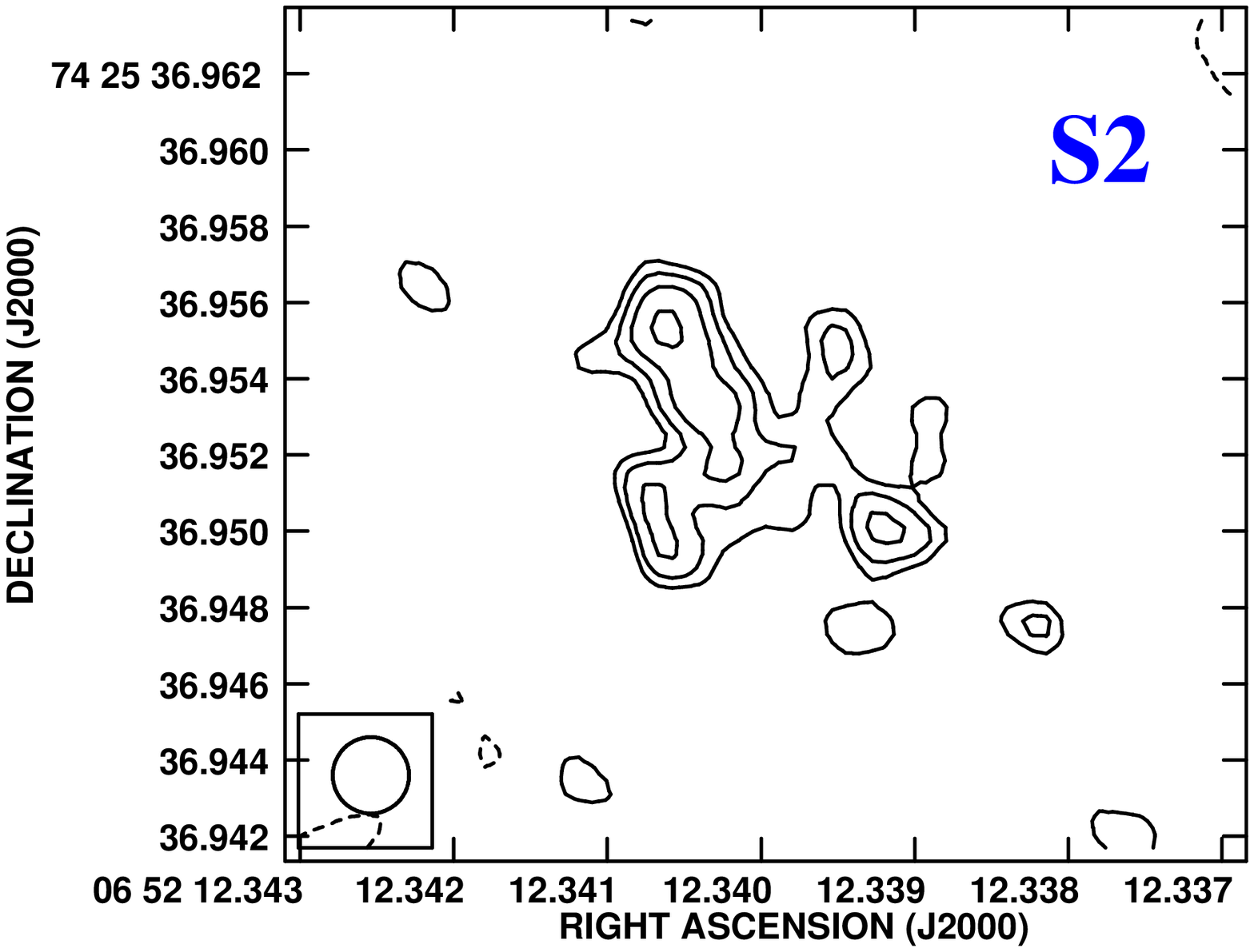}}
\caption{The 4.9~GHz radio image of Mrk\,6: only components S1 (top) and S2 (bottom) are detected. Component S1 seems to have two peaks, although there seems to be more sub-structure. The contours are in percentage of peak intensity and increase in steps of $\sqrt{2}$. The peak intensity is 0.7 mJy~beam$^{-1}$, and the lowest contour levels are for (S1) $\pm$32\% and (S2) $\pm$22.5\%. The images were convolved with circular beams of size 2~mas\,$\times$\,2~mas.}
\label{figcband}
\end{figure}

We detected three distinct radio components at 1.6~GHz. These are identified as N, S1 and S2 in Figure~\ref{figwhole}. Components S2, S1, and N correspond to components 2, 4 and 6 of Figure~1 of \citet{Kukula96}, respectively. All components appear resolved: the position angles of their extensions are indicated in the left panel of Figure~\ref{figwhole}. 
{The centre of the optical host galaxy of Mrk\,6 (indicated as a cross in Figure~\ref{figwhole}) is at RA}
06h~52m~12.251s, Dec +74$\degr$~25$\arcmin$~37.46$\arcsec$, with an uncertainty ellipse of 0.36$\arcsec\times0.30\arcsec$ at PA 90$\degr$ \citep{Li96}. At 4.9~GHz, we detected components S1 and S2 {only} (see Figure~\ref{figcband}). 

\begin{figure}
\centering{\includegraphics[width=10cm]{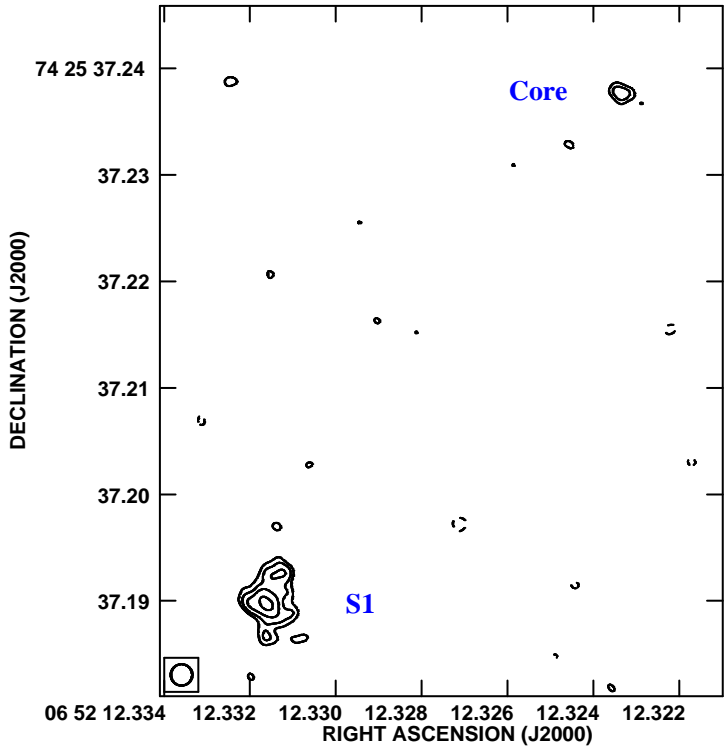}}
\caption{The 4.9~GHz radio image of Mrk\,6 showing the radio core. The image is convolved with a circular beam of size 2~mas\,$\times$\,2~mas. The contours are in percentage of peak intensity and increase in steps of $\sqrt{2}$. The peak intensity is 0.7 mJy~beam$^{-1}$, and the lowest contour levels are $\pm$32\%. The distance between the core and component S1 is 0$\farcs$058 or $\approx$22 pc.}
\label{figccore}
\end{figure}

We also detected with a significance of 6$\sigma$ a compact component at 4.9~GHz which {we propose is} the radio core of Mrk\,6, at a position of RA 06h~52m~12.32333$\pm$0.00004s, Dec 74$\degr$~25$\arcmin$~37.2376$\pm$0.0002$\arcsec$ (see Figure~\ref{figccore}). {The identification of this compact component as the radio core is validated by its spectral index and brightness temperature values, as discussed below.} {We note that the systematic errors in the component positions are much larger than the errors quoted here from JMFIT, and are typically of the order of 0.1$-$0.2 mas \citep[e.g.,][]{Ma98}. The peak intensity and total flux density of the core is 0.4\,mJy\,beam$^{-1}$ and 0.5\,mJy, respectively.} The beam-deconvolved size of the core is 1.86~mas\,$\times$\,0.46~mas, (=0.7~pc\,$\times$\,0.2~pc; however see more discussion on this below). The core minor axis of 0.2~pc corresponds to $10^4$ times the Schwarzschild radius ($R_s$) in Mrk\,6, for a black hole mass of $\sim2\times10^8$M$_\sun$ \citep{Doroshenko12}. {While compact jet-dominated cores are expected to have sizes of $\sim10^3-10^4\,R_s$ (similar to what we find in Mrk\,6), and advection-dominated accretion flows \citep[ADAF;][]{Mahadevan97} should have core sizes of $\sim$10s of $R_s$ \citep{Ulvestad03,Anderson05}, the highest achievable angular resolution on Earth restricts one from truly differentiating between a jet-dominated and an ADAF-dominated core. }

We note that the core {detected here} is distinct from component 5 of Figure~1 of \citet{Kukula96}, as it is only 58~{mas} (=22 pc) from component S1, while component 5 is about 200~{mas} (=75 pc) away from component 2 (i.e., component S1 in present paper). Therefore, VLBI observations were crucial in the identification of the radio core in this galaxy. Component 5 of Kukula et al., which was not detected in the present {data}, could be counterjet emission, as the real core lies between component 5 and component 4 (S1 here). This counterjet emission must be more diffuse than the jet emission (S1), so that VLBI observations fail to detect it.

The mean spectral index of component S1 is $-0.3\pm0.2$, and component S2 is $-0.6\pm0.2$. The relatively flat spectral index of component S1 could be suggestive of the jet segments being currently powered by the AGN {(i.e., with relativistic particles being currently injected into it)}, or to particle acceleration in a shock. As the core was not clearly detected either at 1.6 GHz or 8.4 GHz, we use {two times the rms noise values in the region of the expected core as an upper limit to the core flux density at 1.6~GHz and 8.4~GHz. For an integrated core flux density at 4.9~GHz of $\sim$0.41~mJy, we estimate that the 1.6$-$4.9~GHz and 4.9$-$8.4~GHz spectral indices are $\ge0.4$ and $\le-0.05$, respectively. Two times the rms noise was chosen as a conservative upper limit because of the dependence of noise on the number of CLEAN iterations specified in IMAGR. These spectral index limits are in general agreement with the flat or inverted radio spectra routinely observed in Seyferts cores on different spatial scales \citep[e.g.,][]{Nagar00,Ulvestad01}.}

\begin{table*}
\caption{Observed and Derived Parameters}
\begin{tabular}{lllllllllll}
\hline\hline
{Comp}&{Flux}&{Length}&{Width}&{B$_{min}$}&{P$_{min}$}&{t$_{life}$}&{T$_{\mathrm B}$}\\
{}&{mJy}&{mas}&{mas}&{milliG}&{dyn\,cm$^{-2}$}&{yr}&{K}\\
\hline
S1\_1.6  & 2.9 & 12.1 & 4.7 & 1.2 & 4.6$\times10^{-8}$ & 8772 & 3.4$\times10^{7}$ \\
S2\_1.6  & 3.8 & 12.6 & 10.2& 1.0 & 2.9$\times10^{-8}$ & 11531& 2.0$\times10^{7}$ \\
N\_1.6  & 5.8 &  31.5 & 19.7 & 0.6 &1.0$\times10^{-8}$ &1172 & 6.2$\times10^{6}$ \\
S1\_4.9  & 2.3 & 7.0  & 5.7 & 1.7 & 9.6$\times10^{-8}$ & 5202 &4.4$\times10^{6}$   \\
S2\_4.9  & 2.0 & 7.9  & 6.9 & 1.4 & 5.9$\times10^{-8}$ & 6961 & 2.7$\times10^{6}$  \\
\hline 
\end{tabular}\\
{Column~1: Component name: S1\_1.6 and S1\_4.9 refer to component S1 at 1.6 and 4.9~GHz, respectively, {and so on}. Columns~2, 3 and 4: The integrated flux {density}, length and width of the resolved components were obtained from the elliptical Gaussian-fitting task JMFIT in AIPS. {These estimates were derived from 1.6 and 4.9 GHz images convolved with circular beams of size 5~mas. As component N was resolved out in the 5~mas image, a 6~mas image was used instead for this component.} Columns~5 and 6: Minimum energy magnetic field strength and pressure, respectively. {The spectral index for component N was assumed to be the same as component S2 =$-0.6$.}  Column~7: Lifetime of electrons undergoing synchrotron radiative and inverse Compton losses on CMB photons for a break frequency of 4.9~GHz. Column~8: Brightness temperature {at 1.6 and 4.9~GHz, respectively, for} the resolved components.}
\label{tabprop}
\end{table*}

We estimated the brightness temperature of the radio components using the relation, $T_{B}=1.8\times10^{9}~(1+z)~(\frac{S_{\nu}}{1~mJy})~(\frac{\nu}{1~GHz})^{-2}~(\frac{\theta_{1}\theta_{2}}{mas^{2}})^{-1}~K$, where $\theta_{1}, \theta_{2}$ are the major, minor axes of the best-fitting elliptical Gaussian for a resolved radio component \citep{Ulvestad05}. Component sizes were derived using the Gaussian-fitting task JMFIT in AIPS. For the unresolved core, we multiplied ($\theta_{1}\theta_{2}$) by a factor of 1/2, under the assumption that half the beamsize was the upper limit to the size of the unresolved component \citep[see][]{Ulvestad05}. The brightness temperature of the radio core is $1\times10^8$~K, and is consistent with {non-thermal synchrotron emission. The brightness temperature is a lower limit because the assumed core size is an upper limit.}
$T_{B}$ is typically around $\sim10^7$~K for components S1, S2 and N (see Table~\ref{tabprop}), and is similar to the brightness temperatures observed in other Seyfert nuclei \citep[e.g.,][]{Kukula99,Middelberg04,Orienti10}. The high brightness temperatures provide support for a non-thermal origin of the radio emission. 

Component S1 resembles a jet element at a PA of $\sim150\degr$, while components S2 and N are arc-like or hot spot-like features that are stretched in a direction roughly perpendicular to the PA of the jet feature in component S1. The PA of a line joining the core and S1 is $\approx145\degr$, close to the elongation of component S1 itself. The bright edges and the hot spots along the edges of the kiloparsec radio lobes, resemble the projection of a highly curved radio jet. 

We estimated the magnetic field strength for a cylindrical jet geometry, under the minimum-energy assumption for relativistic particles and magnetic fields \citep{Burbidge59,Miley80}. Images convolved with 5~mas beams were used to obtain the flux densities and sizes of the radio components (with JMFIT). The minimum-energy magnetic field strength was obtained by numerically minimising the sum of electron and magnetic field energy density, assuming powerlaw electron energy spectra with energy indices corresponding to $\alpha=-0.3$ for component S1 and $\alpha=-0.6$ for component S2. We assumed that for all the components $\gamma_{\rm min}=10$, $\gamma_{\rm max}=10^5$ (the results depend only weakly on these assumptions). For a true minimum energy density, we assumed no additional contribution to the energy density from protons and a filling factor of unity. The minimization was carried out using the code of \citet{Hardcastle98}. The minimum $B$-field strength turns out to be of the order of a milliG, while the minimum pressure is a few times $10^{-8}$ dyne cm$^{-2}$ (see Table~\ref{tabprop}).

As the radio core is detected at 4.9~GHz, but not at 1.6 GHz or 8.4~GHz, we can assume the turnover frequency ($\nu_m$) of the core's powerlaw emission to lie between 1.6 and 8.4~GHz. Then an estimate of the magnetic field strength (in G) can be made, assuming that the frequency turnover is due to synchrotron self-absorption rather than free-free absorption, following the relation from \citet{Kellermann81}: $B=\frac{\nu_m^5}{f(\gamma)^5 S_m^2 \theta^{-4} (1+z)}$, where $f(\gamma)$ is a weakly dependent function of the electron energy index $\gamma$, and is $\sim$8 for $\gamma$=2 ({\it i.e.,} for $\alpha=\frac{1-\gamma}{2}=-0.5$); $S_m$ is the maximum synchrotron flux density in Jy, $\nu_m$ is in GHz, and $\theta$ is the angular size in milliarcseconds. As the derived magnetic field strength is a very strongly dependent function of $\nu_m$ and $\theta$, and the Gaussian-fitting task JMFIT typically fails to constrain the minor axis of the deconvolved core, there can be a large uncertainty in the derived magnetic field strength. For an assumed $\nu_m$ of 5 GHz, the $B$ field can vary from $1.5\times10^5$ G to {2~mG}, for $\theta$ varying from [1.86~mas\,$\times$\,0.46~mas, or 0.7~pc\,$\times$\,0.2~pc] to {[1.86~mas\,$\times$\,0.005~mas, or 0.7~pc\,$\times$\,0.002~pc]}. Note that the former $\theta$ is derived from one set of ``nominal'' major and minor axes values derived with JMFIT, the minor axis in the latter $\theta$ has been chosen to be arbitrarily small (since JMFIT does not typically constrain it but only indicates that it is much smaller than the beam-size of the image = 2~mas\,$\times$\,2~mas). This latter $\theta$ value was in fact chosen to get a $B$ field estimate that was comparable to the minimum-energy magnetic field value derived in Table~\ref{tabprop}. However, if the turnover frequency was closer to 1.6 GHz, then the unconstrained minor axis of the deconvolved core {could be 4 times larger and the $B$ field could still be close to equipartition ($\sim2$ mG). All in all, our data are not precise enough to get a robust magnetic field estimate, and we cannot say with certainty if the $B$ field is close (or not) to the equipartition value.} \citet{Bontempi12} have suggested that the excessively large magnetic field estimates derived in their VLBI observations of Seyfert cores are consistent with thermal, rather than synchrotron emission.

The lifetime of electrons in the jet components undergoing both synchrotron radiative and inverse Compton losses on CMB photons, was estimated using the relation $t_{e}\simeq2.6\times10^{4} \frac{B^{1/2}}{B^{2}+B_{r}^{2}} \frac{1}{[(1+z)\nu_{b}]^{1/2}}$ years \citep{vanderlaan69}, where $B$ was assumed to be the equipartition magnetic field $B_{min}$, and the break frequency $\nu_{b}$ was assumed to be 4.9~GHz. 
{The magnetic field equivalent to the radiation, which was assumed to be predominantly CMB photons, $B_{r}$,} was estimated using the relation, $B_{r}\simeq4\times10^{-6} (1+z)^{2}$ Gauss, where $z$ is the source redshift. The electron lifetimes lie between 5000$-$11,000 years. These lifetimes are about 30 percent shorter than if only synchrotron losses were prevalent. However, it has been pointed out that stellar photon density may be much higher than the CMB photon densities in nearby elliptical galaxies \citep{Stawarz03,Hardcastle11}. As this may also hold for bulge-dominated S0 type galaxies which host Seyfert nuclei, we can expect that the inverse Compton losses will be higher than estimated here, and this will reduce the electron lifetimes further. An additional caveat is that since self-calibration could not be carried out while imaging the source, a significant amount of flux density could be missing from the radio components. A higher radio flux density will imply higher $B_{min}$ values and lower electron radiative lifetimes. 

\section{Discussion}
\begin{figure}
\centering{\includegraphics[width=12cm]{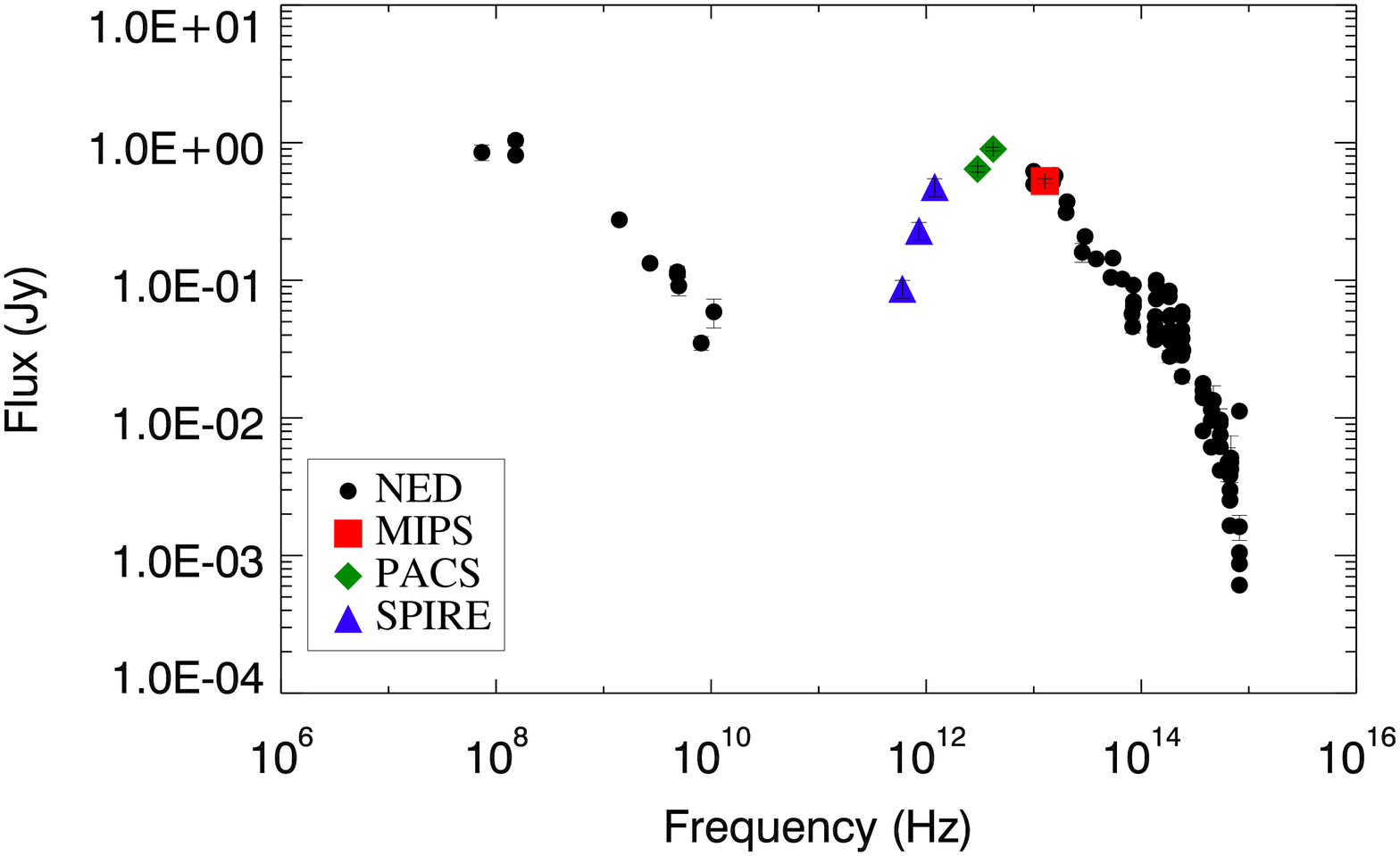}}
\caption{Broad-band SED of Mrk\,6 with new infrared data from the {\it Herschel} space telescope. Other data are from NED.}
\label{figsed}
\end{figure}

In \citet{Kharb06}, we had estimated that the star formation rate derived from {\it Spitzer} observations was far below that required to inflate the radio bubbles. Additional evidence for a comparatively low SFR comes from new {\it Herschel} data which we obtained from the {\it Herschel} archive and reduced using the {\it Herschel} analysis software HIPE (see Section~2 for details). We present in Figure~\ref{figsed} the broad-band SED of Mrk\,6 with new {\it Herschel} infrared data using the PACS and SPIRE instruments \citep{Poglitsch10,Griffin10}. {No previous data existed in this part of the spectrum and this figure clearly shows the importance of the new {\it Herschel} data for tracing the infrared peak of the SED, which originates in AGN and/or star formation heated dust emitting in the infrared.} These {\it Herschel} data have the advantage over previous {\it Spitzer} measurements in that the star formation is thought to be the dominant heating mechanism of the dust that produces emission at these longer wavelength in the far-infrared. At {\it Spitzer} wavelengths (i.e. 24$\mu$m) a considerable contribution to the 24$\mu$m luminosity may be expected from the AGN instead. Therefore, the SPIRE 250$\mu$m band can give a better estimate of the SFR than was previously possible with {\it Spitzer}. Calculating {the luminosity at 250$\mu$m (L250$\mu$m)} for Mrk\,6 and using the empirical relation derived by \citet{Hardcastle13} for L250$\mu$m versus SFR, measured from optical spectral modelling, we find a low SFR of 0.8~M$_\sun$ yr$^{-1}$. This SFR is also an upper limit because the 250$\mu$m flux can originate both from cool dust heated by the old stellar population as well as from young star forming regions. It is also noteworthy that Mrk\,6 has no strong evidence for star formation tracing polycyclic aromatic hydrocarbon (PAH) features in contrast to the majority of Seyfert galaxies \citep{Gallimore10}. Therefore, for the remaining discussion we have assumed that AGN jets are primarily responsible for the creation of the radio bubbles. We discuss the morphology of the radio jets, their connection to the kpc-scale bubbles, and the bubble ages in the following sections. Finally we discuss the radio-loudness of Mrk\,6 and other Seyfert galaxies and compare them to those of low luminosity FRI radio galaxies.

\subsection{Changes in Jet Direction: Interaction with the Medium}
\citet{Kinney00} used the distribution of radio jet orientation in Seyfert galaxies to show that the accretion disks are commonly misaligned with the host galaxy disk. There are several reasons why this might be the case, including misaligned inflow of gas from a minor merger. The misaligned gas inflow will change the accretion disk axis, and the blackhole will re-align with the disk via Lense-Thirring precession \citep{Bardeen75,Rees78}. Minor mergers could be responsible for the creation of randomly-oriented short-lived accretion disks in Seyfert galaxies, that could in turn lead to jet precession \citep[e.g.,][]{Volonteri07}. Another mechanism for a change in the jet axis would be a merger with another black hole \citep[e.g.,][]{Merritt02}. {For the brightest cluster galaxy of RBS797, which exhibits orthogonal radio outflows, a jet flip due to the presence of a binary black hole has been suggested by \citet{Gitti13}.} ISM rotation and ram pressure bending have also been invoked to explain the curved radio jets in some Seyfert galaxies \citep[e.g.,][]{Wilson82}. While jet flips alone may arise from a minor merger event, both jet precession and jet flips, as discussed for Mrk\,6 ahead, can naturally occur due to the presence of binary black holes. 

\begin{figure}
\centering{\includegraphics[width=13cm]{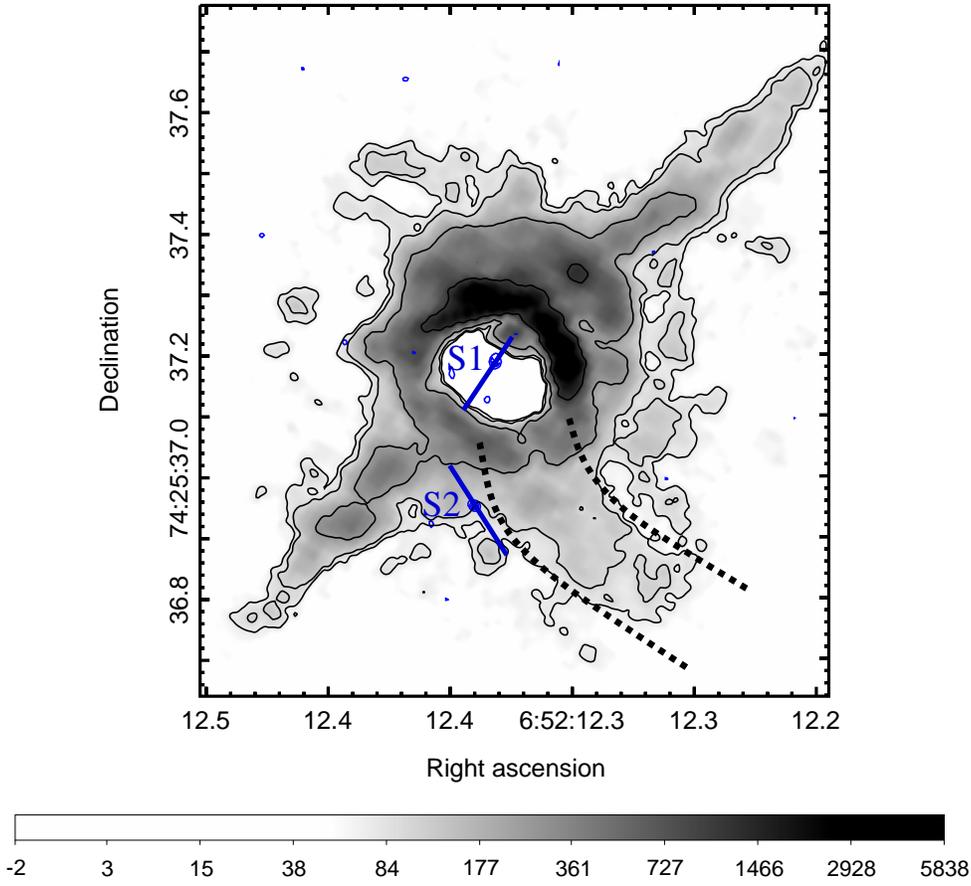}}
\caption{The {\it HST} archival FOC/F372M image showing the emission line gas structure highlighted by the black dashed lines in Mrk\,6. The blue contours denote the VLBA emission. The PA of components S1 and S2 are represented by blue solid lines. Component S2 seems to lie at the outer edge of the curved emission line structure.}
\label{figfoc}
\end{figure}

The archival {\it Hubble Space Telescope (HST)} Advanced Camera for Surveys (ACS) F330W image\footnote{Program ID 9379, PI - Henrique Schmitt, http://archive.stsci.edu/cgi-bin/mastpreview?mission=hst$\&$dataid=J8EK17011} of Mrk\,6 shows the presence of a tail-like feature towards the north-west, which could be a signature of past merger activity in Mrk\,6. The emission line images of [OIII] and [OII] from the Faint Object Camera (FOC) aboard the {\it HST} show that there is dense emission line gas in the direction of the nucleus in Mrk\,6, which could provide the clouds which merge with the accretion disk. The emission line images show a clear correspondence with the southern part of the MERLIN radio jet \citep[see][]{Capetti95}. Capetti et al. however used a weak radio component in between S1 and S2 as the radio core to align their emission-line image. 
{When the detected radio core is properly aligned, we find that the radio component S2 lies at the outer edge of the bend observed in the S-shaped emission line region (see Figure~\ref{figfoc}).}
This close correspondence could be indicative of interaction between the radio ejecta and the surrounding medium \citep[e.g.,][]{Capetti95}. {\it Chandra} and {\it XMM-Newton} observations of Mrk\,6 have shown that the X-ray emitting gas has a complex structure \citep{Schurch06,Mingo11}. This is also consistent with a jet that is changing its direction of propagation, interacting with the surrounding ISM. 

\subsection{A Curved Jet due to Precession ?}
\label{secprec}
We first examine the precession model to explain the S-shaped jets in Mrk\,6. The precession model is in keeping with the basic idea proposed by \citet{Kukula96}, namely, that the north-south jets are producing the hot spots along the edges of the kpc-scale north-south bubbles. We note that Kelvin-Helmholtz instabilities that arise in relativistic jets due to some perturbation at the jet nozzles could also produce helical twisting patterns in jets \citep[e.g.,][]{Hardee87}. At the end of the section, we examine an alternate model proposed by \citet{Mingo11} to explain the Mrk\,6 lobes.

\begin{figure}
\centering{\includegraphics[width=10cm]{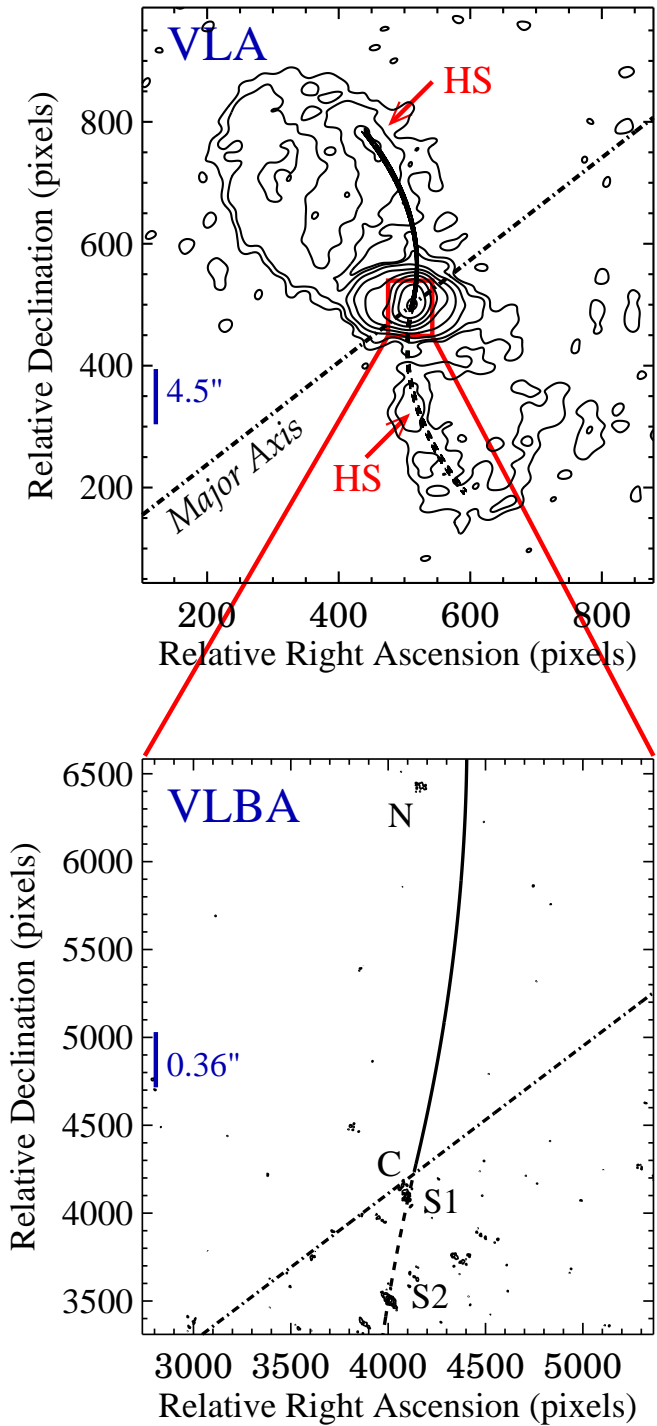}}
\caption{A precessing jet model shown in solid (representing jet) and dashed (representing counter-jet) lines connects the VLBA components (bottom panel, 1 pixel = 1.2 mas) to the hot spots, HS, in the VLA image (top panel, 1 pixel = 45 mas) of Mrk\,6. The 20cm VLA image is reproduced from \citet{Kharb06}. The dot-dashed line indicates the major axis of the host galaxy bulge. The best fit model parameters are discussed in Section~\ref{secprec}}
\label{figprec}
\end{figure}

Following the relations of \citet{Hjellming81}, we created a precessing jet model, and superimposed it on the VLBA and VLA radio images of Mrk\,6 (see Figure~\ref{figprec}). {The 20cm VLA image is reproduced from \citet{Kharb06}.} The new core position was chosen as the origin for this jet model. The projected precessing jet model matches the radio features well. The best fit model parameters {for a clockwise rotating jet} are: jet speed = 0.13$c$ ($\sim3.9\times10^4$~km~s$^{-1}$), jet PA = $-87\degr$, jet inclination = 144$\degr$, jet precession cone half-opening angle = 37$\degr$, and jet angular velocity = 1.5$\times10^{-6}$~rad~yr$^{-1}$. It is important to note that the above set of parameters is not unique, and another combination of the parameter set could also result in a good fit to the apparent radio morphology. This is also demonstrated by the fact that the best fit parameters presented here differ from those derived in \citet{Kharb06}, although different spatial scales were examined in the two cases. One therefore needs independent measurements of as many parameters as possible (e.g., jet speed, jet inclination), to realistically constrain the precession model. However, it is important to note that precessing jets have independently been inferred in Seyfert galaxies, through core flux and PA variability studies, carried out by multi-epoch VLBI obervations \citep[e.g., III\,Zw\,2 \& M\,82;][]{Brunthaler05,MartiVidal11}.

For the best fit jet speed and inclination obtained here, a jet-to-counterjet intensity ratio ($\equiv[1+\beta\cos\theta/1-\beta\cos\theta]^p$) {of 0.6 is expected for a jet structural parameter, $p$ of 2.3, and 0.5 for a $p$ of 3.3 \citep[$p=2-\alpha$, or $3-\alpha$ for a continuous or ``blobby'' jet, respectively;][]{UrryPadovani95}.} It is interesting though that no jet or counterjet emission is detected to within 22 parsec of the core. While such a gap is traditionally considered to be a signature of relativistic deboosting in fast jets that are lying close to the plane of the sky, it is difficult to assume the same for Seyfert galaxies, since Seyfert jets are expected to be only mildly relativistic \citep[e.g.,][]{Bicknell98,Middelberg04}.

{The way it is} depicted in Figure~\ref{figprec}, the jet has been ``on'' for 1.4$\times10^5$ yr. This time constraint follows from the basic idea proposed by \citet{Kukula96} that the north-south jet gets disrupted after forming hot spots along the edges of the north-south lobes. Therefore, 1.4$\times10^5$ yr was the time the jet must have been ``on" in the precession model, in order for it to be disrupted around the position of the bright edges or hot spots of the north-south bubbles {(as determined visually from Figure~\ref{figprec})}.
From this best fit model, it {is apparent} that component N is not a part of the jet that connects the parsec scale emission to the kpc-scale north-south bubbles, but is rather a part of the east-west bubbles. This is also seen by the PA of its extension in the left panel in Figure~\ref{figwhole}. The east-west extensions in Figure~3 of \citet{Kukula96}, which are clearly part of the east-west bubbles, also have the same PA. The fact that this radio component has the steepest spectra (actually only a lower limit) of the three detected components, could be consistent with it arising from a different AGN activity episode, compared to components S1 and S2. The steep spectrum of component N could suggest electron radiative decay in older radio emission. However, this is inadequate evidence by itself, since the spectrum of component S2 is also steep. {While the MERLIN observations of Kukula et al., were unable to resolve components S1 and S2 for the purpose of estimating the 1.6$-$4.9~GHz spectral indices, they derived an average $\alpha=-0.8$ for that part of the jet. Their component 6 (component N here) was easily resolved and had a somewhat steeper spectral index of $-0.9$.}

Apart from the redundancies present in the best-fitting model, another caveat of the precession model, as presented here, is that it suggests that the jet is currently powering the larger north-south lobes, rather than the smaller but brighter east-west lobes. While the extended narrow line region, which is likely being photoionized and/or shock-ionized by the AGN, does lie along the north-south direction, consistent with this being the current jet direction; the size and surface brightness of the east-west lobes are suggestive of them being the {ones that are newly formed.}

Another model, suggested by \citet{Mingo11}, proposes that the bubbles are edge-brightened lobes created by the AGN that are expanding into the ISM. This scenario also requires a flip in the AGN {jet} axis. This model is consistent with the X-ray observations which detect shocked gas (albeit only in small regions) ahead of the radio lobes. A drawback of this model is that it cannot explain why the upper edge of the northern lobe and the lower edge of the southern bubble is relatively brighter than the opposite edge. This effect is much more pronounced in the east-west bubbles where the lower edge of the east bubble and upper edge of the west bubble is brighter. If we assume that the ram pressure of the lobe against a counter-clockwise rotating ISM is responsible for the selective edge brightening in the north-south lobes, then the ISM rotation would need to be clockwise to explain the bright opposite edges in the east-west lobes. This could be a big drawback for the edge-brightened lobe model. Selective edge brightening can more easily be explained as directional, {albeit} curved, AGN outflows. 

\subsection{Constraints on Bubble Ages from Radio and X-ray Observations}
\citet{Mingo11} estimated that the total work available from the plasma filling the north-south lobes was $\approx4\times10^{56}$~erg. For the AGN jet with a total radio luminosity of say, $\approx10^{40}$~erg~s$^{-1}$ {\citep[][Table~3]{Kharb06},} and kinetic luminosity of $10^{42}$~erg~s$^{-1}$ \citep[assuming an efficiency of conversion of 1 percent;][]{Odea85,Birzan07}, the jet would require $1.3\times10^{7}$ years to power the north-south lobes. This is about 90 times longer than the time the jet was ``on'' in the precessing jet model. The two timescales would be comparable if the conversion efficiency was lower than 0.1 percent.  

Based on the Mach number $M$ of the shock producing the X-ray emission at the outer edges of the north-south bubbles in Mrk\,6 ($M=3.9$), \citet{Mingo11} estimated that the speed of the plasma in the bubbles was around 1030$^{+75}_{-100}$ km~s$^{-1}$. From the {\it Chandra} observations, they also estimated that the electron density of the ISM outside the bubbles varied from 8 - 40$\times10^{-4}$~cm$^{-3}$. This implies that the ISM gas mass density {(obtained by multiplying the ISM electron density with the mass of the hydrogen atom = $1.67\times10^{-24}$ g)} varies from 1.3$-$6.7$\times10^{-27}$~g~cm$^{-3}$. If the minimum pressure of the north bubble varies from $1.2\times10^{-12}$~dyne~cm$^{-2}$ to $6.2\times10^{-11}$~dyne~cm$^{-2}$ for a filling factor of 1.0 and $10^{-3}$, respectively \citep[see Table 3,][]{Kharb06}, then the velocity of the lobe plasma can be estimated assuming that the lobe is in ram pressure balance with the surrounding ISM. Using the relation $P_{min} = \rho_{ISM} v_{lobe}^2$ and the ISM gas density from above, we find that $v_{lobe}$ varies between $\sim$130 to 960 km~s$^{-1}$ for a gas density of 6.7$\times10^{-27}$~g~cm$^{-3}$, and between $\sim$300 to 2150 km~s$^{-1}$ for a gas density of 1.3$\times10^{-27}$~g~cm$^{-3}$. Assuming that the plasma velocity inside the lobes was around 1000~km~s$^{-1}$, the north bubble of extent 7.7 kpc would take 7.5$\times10^6$ years to form, while the east bubble of extent 1.4 kpc would take 1.4$\times10^6$~yr to form for the same plasma speed. However, as {\it Chandra} could not clearly resolve the X-ray emission from the east-west bubbles, the plasma speed and therefore the age is uncertain for the east-west structure. It is only clear that the jet axis in Mrk\,6 has changed at least once within the last $\approx10^7$~yr. 

\subsection{The Radio Loudness of Seyferts}
Seyfert galaxies have been historically identified as radio-quiet AGN.
\citet{Kellermann89} defined the radio-loudness parameter, $R$, of an AGN to be the ratio of its radio flux density at 5~GHz to its $B$-band optical flux density at 4400 \AA, with the radio-quiet/radio-loud AGN division set at $R\sim10$. {As Seyferts reside in spiral or S0-type host galaxies, and radio-loud AGN are typically hosted by elliptical galaxies, explanations for the ``lack" of large-scale radio emission in Seyferts, have focused on differences in their central engines, which are in turn dependent on their host galaxy types. {For} instance, differences in black hole masses and spins, both of which are different in spiral and elliptical galaxies, have been suggested to explain the differences in the radio-loudness properties \citep[e.g.,][]{Sikora07}.} It is in this context that it is interesting to note that Mrk\,6 has an $R$ value of $\approx23$, making it a radio-loud Seyfert galaxy (the total 5 GHz flux density as measured by the single dish Green Bank Telescope for Mrk\,6 is 115$\pm$9 mJy \citep{Gregory91}, while the $B$-band apparent magnitude is 14.88$\pm$0.16 mag (NED)\footnote{NASA/IPAC Extragalactic Database}). The mere existence of radio-loud Seyferts can provide strong constraints on the jet production mechanisms in AGN.

The inherent assumption in the estimation of $R$ above, is that the $B$-band luminosity is dominated by the AGN light, which is unlikely to be true for Seyfert galaxies. 
\citet{HoPeng01} extracted the optical nuclear luminosities for a large number of Seyfert galaxies through a proper modelling of their galactic bulge emission in {\it HST} images. Remarkably, after taking into account their optical nuclear luminosities, Ho \& Peng found that a majority of Seyfert 1 galaxies were radio-loud. A similar conclusion was reached by \citet{Terashima03} on the basis of the radio to X-ray flux density ratio for a sample of low luminosity AGN including Seyferts. We present in the Appendix a comparison of the radio-loudness parameters of Seyfert (1 and 2) galaxies belonging to the Extended 12$\mu$m sample of \citet{Rush93}, along with a sample of UGC FRI radio galaxies, having taken into account their optical nuclear luminosities. We estimate that, even without {acknowledging} the AGN-only contribution to the optical luminosity, $\approx$11 percent of Seyfert galaxies (both type 1s and 2s) from the Extended 12$\mu$m complete sample of \citet{Rush93} can be classified as radio-loud following the Kellermann et al. definition. This fraction increases to $\ge$70 percent, when the optical nuclear luminosities are considered (see Table~\ref{tabradl}). 

\begin{figure}
\centering{\includegraphics[width=10cm]{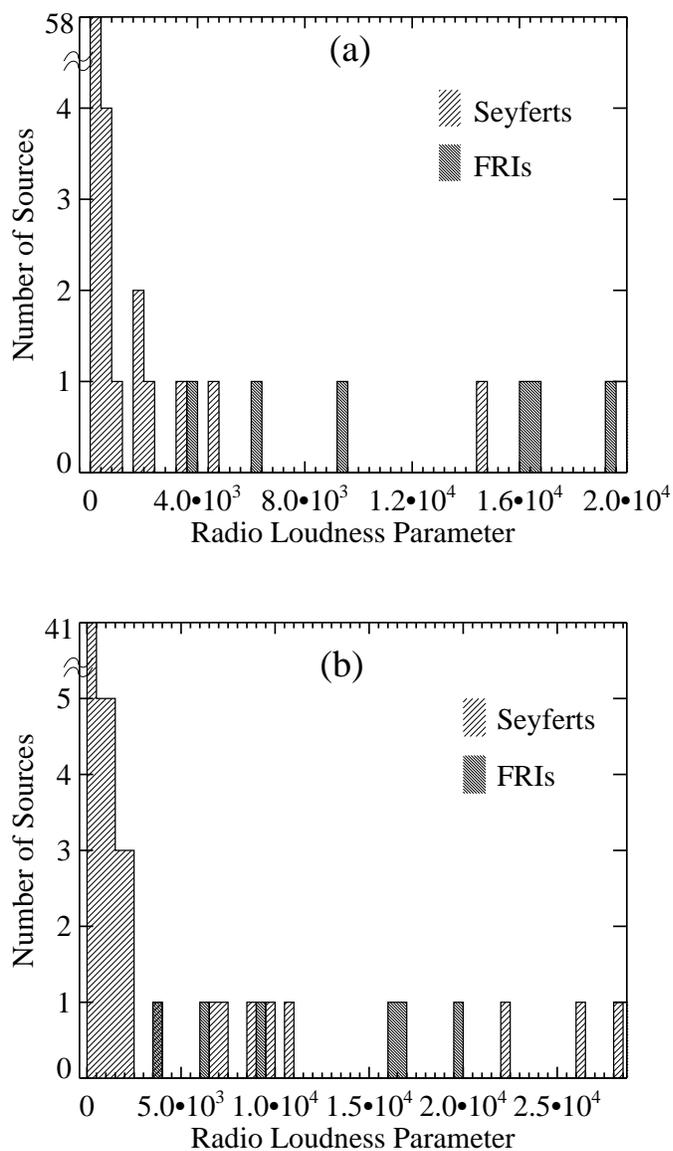}}
\caption{The radio-loudness parameter ($R$) for Seyfert galaxies belonging to the Extended 12$\mu$m sample \citep{Rush93} and FRI radio galaxies belonging to the UGC sample \citep{Verdoes99}. For the estimation of $R$ for the Seyfert galaxies, median nuclear absolute $B$-band magnitudes of the CfA (=$-$17.4) and the Palomar Seyfert samples \citep[=$-$14.6;][]{HoPeng01} were chosen for panels ``a" (top) and ``b" (bottom), respectively.} 
\label{figrl}
\end{figure}

We have plotted in Figure~\ref{figrl} the radio-loudness parameter values for these Seyfert galaxies, along with those of 21 UGC FRI radio galaxies \citep{Verdoes99,Kharb12a}. There is an overlap between the $R$ values of this subset of Seyferts and those of low luminosity FRI radio galaxies. \citet{Balmaverde06} and \citet{Kharb12b} also found a continuum of properties in the host galaxy surface brightness profiles as observed by the {\it HST}, which reflects on the galaxies' nuclear structures, for a small sample of Seyfert and FRI radio galaxies. Although on average Seyfert galaxies have less massive blackholes than FRI radio galaxies, there is an overlap in the blackhole mass ranges for the two classes: $\sim10^7-10^8$\,M$_\sun$ for Seyferts \citep{Wandel99}, and $\sim10^8-10^9$\,M$_\sun$ for FRI radio galaxies \citep{Woo02}.

\begin{figure}
\centering{\includegraphics[width=9cm]{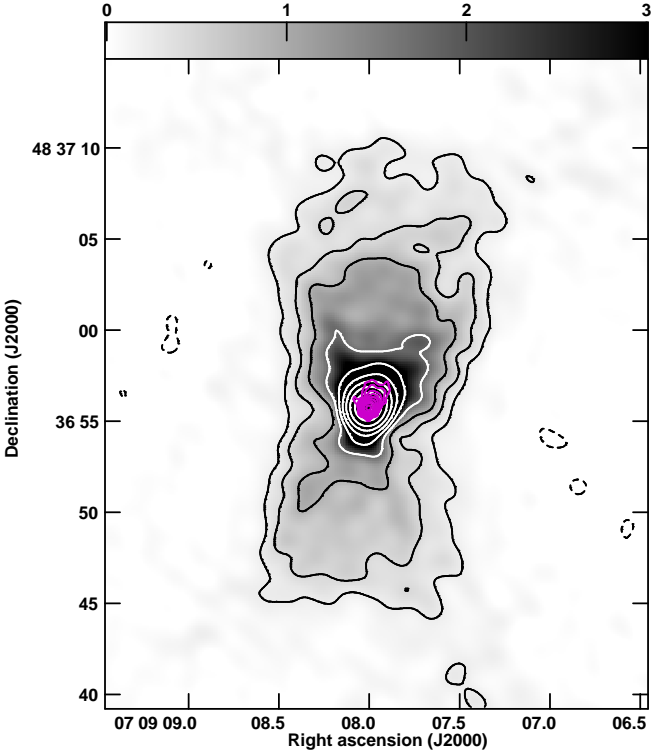}}
\caption{UGC03695: the FRI radio galaxy with the lowest radio loudness parameter in the sample under study. The black and white contours indicate the 1.4 GHz emission (levels are in percentage of peak intensity and increase in steps of 2; the lowest contour is at $\pm$0.17 percent of peak intensity = 128~mJy~beam$^{-1}$), while the magenta contours indicate the 4.9 GHz emission (the lowest contour is at $\pm$0.3 percent of peak intensity = 99~mJy~beam$^{-1}$; Noel-Storr et al., in preparation).}
\label{figrlfr1}
\end{figure}

There is also a continuum in the radio morphological properties of FRI galaxies and radio-loud Seyfert galaxies, although more sensitive observations are required to map the diffuse lobe emission in the latter. In the absence of sensitive radio observations of Seyfert galaxies, we can restrict our comparison to individual sources, which prevents us from claiming a statistically significant overlap in radio morphology between Seyfert galaxies and low luminosity FRIs. The radio structure in the Seyfert galaxy, Circinus, with hot spot like features immersed in bubble-like lobes \citep[see][]{Mingo12}, resembles the hot spot-lobe structure observed in the FRI radio galaxy, Hercules A \citep{Gizani03,ODea13}. While the radio bubbles of Mrk\,6 and other Seyfert galaxies (e.g., NGC\,6764) are typically a few kpc to $\sim$10 kpc across, and are confined to the bulge of the host galaxy, bipolar bubbles of extent $>$100 kpc have been observed in the radio galaxy, NGC\,1167 \citep{Shulevski12}. Interestingly, the FRI radio galaxy with the lowest nuclear radio-loudness parameter in the sample under study, UGC\,03695 ($R$=3775), has a radio structure very similar to that seen in Seyfert galaxies (see Figure~\ref{figrlfr1}). These data are from our VLA program AB920 (J. Noel-Storr et al., in preparation). The lower frequency image (1.4~GHz) shows the bipolar bubble-like lobes, while the higher frequency image (4.9 GHz) shows a jet-like (i.e., elongated) structure pointing toward the western edge of the north lobe. This not only suggests a close connection between the jet-like structure and the bubble-like lobes, but also the presence of a misaligned jet. Small-scale jets that are misaligned with the radio lobes are a common characteristic of Seyfert galaxies.

S0 galaxies resemble elliptical galaxies in terms of their bulge properties, but additionally have low density gas and dust-rich disks in their {centres}. It is {then} plausible that the radio jets in Seyfert galaxies, which likely arise from the inner accretion disks \citep[e.g.,][]{Nixon13} are affected by the presence of these kpc-scale dusty disks, and precess. Any kind of warp that develops in the disk close to the jet launching sites, as a result of misalignment with the large scale galactic disks, may affect the jets so that they are no longer straight. Moreover, the distance from the cores where the jets ``flare'', or decollimate and widen, like in FRI radio galaxies \citep[e.g.,][the flaring distance is typically 1 kpc in FRIs]{Laing02}, could be shorter for Seyfert galaxies. Subsonic jets moving in a confined medium where the density varies more slowly that 1/R$^2$, where R is the distance from the galaxy centre, could produce bubble-like radio structures \citep{Falle91}. Therefore, we propose that a combination of an early-flaring, misaligned jet in a confined medium could produce Mrk\,6-like radio lobes, {and similar lobes observed in other Seyfert galaxies}. 

\section{Summary}
We have presented here high resolution VLBA observations at 1.6 and 4.9 GHz of the $\sim$1~kpc radio jet in the radio-loud Seyfert galaxy, Mrk\,6. 

\begin{enumerate}
\item 
We detect 3 distinct radio components, N, S1, S2, at 1.6 GHz, but only components S1 and S2 at 4.9~GHz. However, the 4.9 GHz observations show the presence of an inverted spectrum radio core ($\alpha^{1.6}_{4.9}$=+1.0$\pm$0.2), with a brightness temperature of $\approx10^8$~K, indicating that it is the base of a relativistic synchrotron jet. This core had not been detected in prior radio observations: VLBI proved to be crucial in the identification of this core. 
\item 
The mean spectral index $\alpha^{1.6}_{4.9}$ of component S1 is $-0.3\pm0.2$ and S2 is $-0.6\pm0.2$. The relatively flatter spectral index of component S1 is consistent with it being currently powered by the AGN, or to particle acceleration in a shock. 
\item 
Additional 8.4~GHz archival VLBA data on Mrk\,6 failed to reveal any radio emission. We estimate that the 4.9$-$8.4~GHz spectral index of the core is steeper than $-0.05$. 
\item 
All the detected components, apart from the core, are resolved and elongated, such that they appear to be either jet segments or hot spots, where the jet impacts the surrounding ISM. The position angles of these components also indicate a connection to one edge of the north-south kpc-scale radio lobes, although no emission is detected continuously to clearly show this connection. More sensitive observations are required to detect the inter-jet-segment emission.
\item 
New {\it Herschel} 250$\mu$m observations provide an upper limit to the star formation rate in Mrk\,6 of 0.8~M$_\sun$ yr$^{-1}$. This estimate is nearly 40 times lower than the SFR required to inflate the radio bubbles 
\citep[$\approx$33~M$_\sun$ yr$^{-1}$;][]{Kharb06}. The AGN jet is therefore primarily responsible for the creation of the radio bubbles in Mrk\,6.
\item 
With the estimated total jet radio luminosity of $\approx10^{40}$~erg~s$^{-1}$, the jet will take $1.3\times10^{7}$~yr to power the north-south bubbles with a total energy of $\approx10^{56}$~erg \citep{Mingo11}. From plasma speeds of the order of 1000~km~s$^{-1}$ obtained through the X-ray observations of \citet{Mingo11}, and ram pressure balance arguments for the ISM and the bubbles, the north-south bubbles are expected to take $7.5\times10^6$~yr to form, and the east-west bubbles $1.4\times10^6$~yr. We suggest that the jet axis has changed at least once in Mrk\,6 within the last $\approx10^7$~yr. This is consistent with the signatures of a minor merger that this galaxy seems to have undergone, as observed in its {\it HST} image. 
\item 
Although from these observations alone, it is difficult to identify the primary mechanism for the curved radio jet in Mrk\,6, jet precession can explain the morphology from parsec to kiloparsec scales. While jet flips alone may arise from a minor merger event, both jet precession and jet flips can naturally occur in the presence of binary black holes.
\item 
Both the {\it HST} imaging which shows a close alignment between the emission-line gas and the radio jet, and the {\it Chandra} X-ray imaging which shows a complex gas structure in the center of Mrk\,6, are consistent with the idea of a jet that is changing its direction of propagation and interacting with the surrounding medium. 
\item 
We find a continuum of radio properties between a subset of radio-loud Seyfert and FRI radio galaxies. 
\end{enumerate}

\section*{Acknowledgments}
{We would like to thank the anonymous referee for suggestions that have improved this manuscript significantly.}
PK would like to thank Mousumi Das for a helpful discussion on galaxy types, and Masanori Nakamura on AGN bubbles. The National Radio Astronomy Observatory is a facility of the National Science Foundation operated under cooperative agreement by Associated Universities, Inc. This research has made use of the NASA/IPAC Extragalactic Database (NED) which is operated by the Jet Propulsion Laboratory, California Institute of Technology, under contract with the National Aeronautics and Space Administration.

\bibliographystyle{mn2e}
\bibliography{ms}

\begin{thebibliography}{77}
\expandafter\ifx\csname natexlab\endcsname\relax\def\natexlab#1{#1}\fi

\bibitem[{{Anderson} \& {Ulvestad}(2005)}]{Anderson05}
{Anderson} J.~M., {Ulvestad} J.~S., 2005, \apj, 627, 674

\bibitem[{{Balmaverde} \& {Capetti}(2006)}]{Balmaverde06}
{Balmaverde} B., {Capetti} A., 2006, \aap, 447, 97

\bibitem[{{Bardeen} \& {Petterson}(1975)}]{Bardeen75}
{Bardeen} J.~M., {Petterson} J.~A., 1975, \apjl, 195, L65

\bibitem[{{Baum} {et~al}\mbox{.}(1993){Baum}, {O'Dea}, {Dallacassa}, {de
  Bruyn}, \& {Pedlar}}]{Baum93}
{Baum} S.~A., {O'Dea} C.~P., {Dallacassa} D., {de Bruyn} A.~G., {Pedlar} A.,
  1993, \apj, 419, 553

\bibitem[{{Beasley} \& {Conway}(1995)}]{BeasleyConway95}
{Beasley} A.~J., {Conway} J.~E., 1995, in Astronomical Society of the Pacific
  Conference Series, Vol.~82, Very Long Baseline Interferometry and the VLBA,
  {J.~A.~Zensus, P.~J.~Diamond, \& P.~J.~Napier}, ed., p. 328

\bibitem[{{Bicknell} {et~al}\mbox{.}(1998){Bicknell}, {Dopita}, {Tsvetanov}, \&
  {Sutherland}}]{Bicknell98}
{Bicknell} G.~V., {Dopita} M.~A., {Tsvetanov} Z.~I., {Sutherland} R.~S., 1998,
  \apj, 495, 680

\bibitem[{{B{\^i}rzan} {et~al}\mbox{.}(2007){B{\^i}rzan}, {McNamara},
  {Carilli}, {Nulsen}, \& {Wise}}]{Birzan07}
{B{\^i}rzan} L., {McNamara} B.~R., {Carilli} C.~L., {Nulsen} P.~E.~J., {Wise}
  M.~W., 2007, in Heating versus Cooling in Galaxies and Clusters of Galaxies,
  {B{\"o}hringer} H., {Pratt} G.~W., {Finoguenov} A., {Schuecker} P., eds., p.
  115

\bibitem[{{Bontempi} {et~al}\mbox{.}(2012){Bontempi}, {Giroletti}, {Panessa},
  {Orienti}, \& {Doi}}]{Bontempi12}
{Bontempi} P., {Giroletti} M., {Panessa} F., {Orienti} M., {Doi} A., 2012,
  \mnras, 426, 588

\bibitem[{{Brunthaler} {et~al}\mbox{.}(2005){Brunthaler}, {Falcke}, {Bower},
  {Aller}, {Aller}, \& {Ter{\"a}sranta}}]{Brunthaler05}
{Brunthaler} A., {Falcke} H., {Bower} G.~C., {Aller} M.~F., {Aller} H.~D.,
  {Ter{\"a}sranta} H., 2005, \aap, 435, 497

\bibitem[{{Buchanan} {et~al}\mbox{.}(2006){Buchanan}, {Gallimore}, {O'Dea},
  {Baum}, {Axon}, {Robinson}, {Elitzur}, \& {Elvis}}]{Buchanan06}
{Buchanan} C.~L., {Gallimore} J.~F., {O'Dea} C.~P., {Baum} S.~A., {Axon} D.~J.,
  {Robinson} A., {Elitzur} M., {Elvis} M., 2006, \aj, 132, 401

\bibitem[{{Burbidge}(1959)}]{Burbidge59}
{Burbidge} G.~R., 1959, \apj, 129, 849

\bibitem[{{Capetti} {et~al}\mbox{.}(1995){Capetti}, {Axon}, {Kukula},
  {Macchetto}, {Pedlar}, {Sparks}, \& {Boksenberg}}]{Capetti95}
{Capetti} A., {Axon} D.~J., {Kukula} M., {Macchetto} F., {Pedlar} A., {Sparks}
  W.~B., {Boksenberg} A., 1995, \apjl, 454, L85

\bibitem[{{Chiaberge}, {Capetti} \& {Celotti}(1999){Chiaberge}, {Capetti}, \&
  {Celotti}}]{Chiaberge99}
{Chiaberge} M., {Capetti} A., {Celotti} A., 1999, \aap, 349, 77

\bibitem[{{Colbert} {et~al}\mbox{.}(1996){Colbert}, {Baum}, {Gallimore},
  {O'Dea}, \& {Christensen}}]{Colbert96}
{Colbert} E.~J.~M., {Baum} S.~A., {Gallimore} J.~F., {O'Dea} C.~P.,
  {Christensen} J.~A., 1996, \apj, 467, 551

\bibitem[{{Doroshenko} {et~al}\mbox{.}(2012){Doroshenko}, {Sergeev},
  {Klimanov}, {Pronik}, \& {Efimov}}]{Doroshenko12}
{Doroshenko} V.~T., {Sergeev} S.~G., {Klimanov} S.~A., {Pronik} V.~I., {Efimov}
  Y.~S., 2012, \mnras, 426, 416

\bibitem[{{Elmouttie} {et~al}\mbox{.}(1998){Elmouttie}, {Haynes}, {Jones},
  {Sadler}, \& {Ehle}}]{Elmouttie98}
{Elmouttie} M., {Haynes} R.~F., {Jones} K.~L., {Sadler} E.~M., {Ehle} M., 1998,
  \mnras, 297, 1202

\bibitem[{{Falle}(1991)}]{Falle91}
{Falle} S.~A.~E.~G., 1991, \mnras, 250, 581

\bibitem[{{Fanaroff} \& {Riley}(1974)}]{Fanaroff74}
{Fanaroff} B.~L., {Riley} J.~M., 1974, \mnras, 167, 31P

\bibitem[{{Gallimore} {et~al}\mbox{.}(2006){Gallimore}, {Axon}, {O'Dea},
  {Baum}, \& {Pedlar}}]{Gallimore06}
{Gallimore} J.~F., {Axon} D.~J., {O'Dea} C.~P., {Baum} S.~A., {Pedlar} A.,
  2006, \aj, 132, 546

\bibitem[{{Gallimore} {et~al}\mbox{.}(2010){Gallimore}, {Yzaguirre},
  {Jakoboski}, {Stevenosky}, {Axon}, {Baum}, {Buchanan}, {Elitzur}, {Elvis},
  {O'Dea}, \& {Robinson}}]{Gallimore10}
{Gallimore} J.~F. {et~al.}, 2010, \apjs, 187, 172

\bibitem[{{Gitti} {et~al}\mbox{.}(2013){Gitti}, {Giroletti}, {Giovannini},
  {Feretti}, \& {Liuzzo}}]{Gitti13}
{Gitti} M., {Giroletti} M., {Giovannini} G., {Feretti} L., {Liuzzo} E., 2013,
  \aap, 557, L14

\bibitem[{{Gizani} \& {Leahy}(2003)}]{Gizani03}
{Gizani} N.~A.~B., {Leahy} J.~P., 2003, \mnras, 342, 399

\bibitem[{{Gregory} \& {Condon}(1991)}]{Gregory91}
{Gregory} P.~C., {Condon} J.~J., 1991, \apjs, 75, 1011

\bibitem[{{Greisen}(2003)}]{Greisen03}
{Greisen} E.~W., 2003, in Astrophysics and Space Science Library, Vol. 285,
  Astrophysics and Space Science Library, {A.~Heck}, ed., p. 109

\bibitem[{{Griffin} {et~al}\mbox{.}(2010){Griffin}, {Abergel}, {Abreu}, \& {176
  coauthors}}]{Griffin10}
{Griffin} M.~J., {Abergel} A., {Abreu} A., {176 coauthors}, 2010, \aap, 518, L3

\bibitem[{{Hardcastle}, {Birkinshaw} \& {Worrall}(1998){Hardcastle},
  {Birkinshaw}, \& {Worrall}}]{Hardcastle98}
{Hardcastle} M.~J., {Birkinshaw} M., {Worrall} D.~M., 1998, \mnras, 294, 615

\bibitem[{{Hardcastle} {et~al}\mbox{.}(2013){Hardcastle}, {Ching}, {Virdee}, \&
  {25 coauthors}}]{Hardcastle13}
{Hardcastle} M.~J., {Ching} J.~H.~Y., {Virdee} J.~S., {25 coauthors}, 2013,
  \mnras, 429, 2407

\bibitem[{{Hardcastle} \& {Croston}(2011)}]{Hardcastle11}
{Hardcastle} M.~J., {Croston} J.~H., 2011, \mnras, 415, 133

\bibitem[{{Hardee}(1987)}]{Hardee87}
{Hardee} P.~E., 1987, \apj, 318, 78

\bibitem[{{Hjellming} \& {Johnston}(1981)}]{Hjellming81}
{Hjellming} R.~M., {Johnston} K.~J., 1981, \apjl, 246, L141

\bibitem[{{Ho} \& {Peng}(2001)}]{HoPeng01}
{Ho} L.~C., {Peng} C.~Y., 2001, \apj, 555, 650

\bibitem[{{Hota} \& {Saikia}(2006)}]{HotaSaikia06}
{Hota} A., {Saikia} D.~J., 2006, \mnras, 371, 945

\bibitem[{{Kellermann} \& {Pauliny-Toth}(1981)}]{Kellermann81}
{Kellermann} K.~I., {Pauliny-Toth} I.~I.~K., 1981, \araa, 19, 373

\bibitem[{{Kellermann} {et~al}\mbox{.}(1989){Kellermann}, {Sramek}, {Schmidt},
  {Shaffer}, \& {Green}}]{Kellermann89}
{Kellermann} K.~I., {Sramek} R., {Schmidt} M., {Shaffer} D.~B., {Green} R.,
  1989, \aj, 98, 1195

\bibitem[{{Kharb} {et~al}\mbox{.}(2012{\natexlab{a}}){Kharb}, {Capetti},
  {Axon}, {Chiaberge}, {Grandi}, {Robinson}, {Giovannini}, {Balmaverde},
  {Macchetto}, \& {Montez}}]{Kharb12b}
{Kharb} P. {et~al.}, 2012{\natexlab{a}}, \aj

\bibitem[{{Kharb} {et~al}\mbox{.}(2010){Kharb}, {Hota}, {Croston},
  {Hardcastle}, {O'Dea}, {Kraft}, {Axon}, \& {Robinson}}]{Kharb10a}
{Kharb} P., {Hota} A., {Croston} J.~H., {Hardcastle} M.~J., {O'Dea} C.~P.,
  {Kraft} R.~P., {Axon} D.~J., {Robinson} A., 2010, \apj, 723, 580

\bibitem[{{Kharb} {et~al}\mbox{.}(2006){Kharb}, {O'Dea}, {Baum}, {Colbert}, \&
  {Xu}}]{Kharb06}
{Kharb} P., {O'Dea} C.~P., {Baum} S.~A., {Colbert} E.~J.~M., {Xu} C., 2006,
  \apj, 652, 177

\bibitem[{{Kharb} {et~al}\mbox{.}(2008){Kharb}, {O'Dea}, {Baum}, {Daly},
  {Mory}, {Donahue}, \& {Guerra}}]{Kharb08}
{Kharb} P., {O'Dea} C.~P., {Baum} S.~A., {Daly} R.~A., {Mory} M.~P., {Donahue}
  M., {Guerra} E.~J., 2008, \apjs, 174, 74

\bibitem[{{Kharb} {et~al}\mbox{.}(2012{\natexlab{b}}){Kharb}, {O'Dea}, {Tilak},
  {Baum}, {Haynes}, {Noel-Storr}, {Fallon}, \& {Christiansen}}]{Kharb12a}
{Kharb} P., {O'Dea} C.~P., {Tilak} A., {Baum} S.~A., {Haynes} E., {Noel-Storr}
  J., {Fallon} C., {Christiansen} K., 2012{\natexlab{b}}, \apj, 754, 1

\bibitem[{{Kinney} {et~al}\mbox{.}(2000){Kinney}, {Schmitt}, {Clarke},
  {Pringle}, {Ulvestad}, \& {Antonucci}}]{Kinney00}
{Kinney} A.~L., {Schmitt} H.~R., {Clarke} C.~J., {Pringle} J.~E., {Ulvestad}
  J.~S., {Antonucci} R.~R.~J., 2000, \apj, 537, 152

\bibitem[{{Kukula} {et~al}\mbox{.}(1999){Kukula}, {Ghosh}, {Pedlar}, \&
  {Schilizzi}}]{Kukula99}
{Kukula} M.~J., {Ghosh} T., {Pedlar} A., {Schilizzi} R.~T., 1999, \apj, 518,
  117

\bibitem[{{Kukula} {et~al}\mbox{.}(1996){Kukula}, {Holloway}, {Pedlar},
  {Meaburn}, {Lopez}, {Axon}, {Schilizzi}, \& {Baum}}]{Kukula96}
{Kukula} M.~J., {Holloway} A.~J., {Pedlar} A., {Meaburn} J., {Lopez} J.~A.,
  {Axon} D.~J., {Schilizzi} R.~T., {Baum} S.~A., 1996, \mnras, 280, 1283

\bibitem[{{Laing} \& {Bridle}(2002)}]{Laing02}
{Laing} R.~A., {Bridle} A.~H., 2002, \mnras, 336, 328

\bibitem[{{Li} \& {Jin}(1996)}]{Li96}
{Li} J., {Jin} W., 1996, \aaps, 120, 201

\bibitem[{{Ma} {et~al}\mbox{.}(1998){Ma}, {Arias}, {Eubanks}, {Fey}, {Gontier},
  {Jacobs}, {Sovers}, {Archinal}, \& {Charlot}}]{Ma98}
{Ma} C. {et~al.}, 1998, \aj, 116, 516

\bibitem[{{Mahadevan}(1997)}]{Mahadevan97}
{Mahadevan} R., 1997, \apj, 477, 585

\bibitem[{{Mart{\'{\i}}-Vidal} {et~al}\mbox{.}(2011){Mart{\'{\i}}-Vidal},
  {Marcaide}, {Alberdi}, {P{\'e}rez-Torres}, {Ros}, \&
  {Guirado}}]{MartiVidal11}
{Mart{\'{\i}}-Vidal} I., {Marcaide} J.~M., {Alberdi} A., {P{\'e}rez-Torres}
  M.~A., {Ros} E., {Guirado} J.~C., 2011, \aap, 533, A111

\bibitem[{{Merritt} \& {Ekers}(2002)}]{Merritt02}
{Merritt} D., {Ekers} R.~D., 2002, Science, 297, 1310

\bibitem[{{Middelberg} {et~al}\mbox{.}(2004){Middelberg}, {Roy}, {Nagar},
  {Krichbaum}, {Norris}, {Wilson}, {Falcke}, {Colbert}, {Witzel}, \&
  {Fricke}}]{Middelberg04}
{Middelberg} E. {et~al.}, 2004, \aap, 417, 925

\bibitem[{{Miley}(1980)}]{Miley80}
{Miley} G., 1980, \araa, 18, 165

\bibitem[{{Mingo} {et~al}\mbox{.}(2011){Mingo}, {Hardcastle}, {Croston},
  {Evans}, {Hota}, {Kharb}, \& {Kraft}}]{Mingo11}
{Mingo} B., {Hardcastle} M.~J., {Croston} J.~H., {Evans} D.~A., {Hota} A.,
  {Kharb} P., {Kraft} R.~P., 2011, \apj, 731, 21

\bibitem[{{Mingo} {et~al}\mbox{.}(2012){Mingo}, {Hardcastle}, {Croston},
  {Evans}, {Kharb}, {Kraft}, \& {Lenc}}]{Mingo12}
{Mingo} B., {Hardcastle} M.~J., {Croston} J.~H., {Evans} D.~A., {Kharb} P.,
  {Kraft} R.~P., {Lenc} E., 2012, \apj, 758, 95

\bibitem[{{Nagar} {et~al}\mbox{.}(2000){Nagar}, {Falcke}, {Wilson}, \&
  {Ho}}]{Nagar00}
{Nagar} N.~M., {Falcke} H., {Wilson} A.~S., {Ho} L.~C., 2000, \apj, 542, 186

\bibitem[{{Napier}(1994)}]{Napier94}
{Napier} P.~J., 1994, in IAU Symposium, Vol. 158, Very High Angular Resolution
  Imaging, {J.~G.~Robertson \& W.~J.~Tango}, ed., p. 117

\bibitem[{{Nixon} \& {King}(2013)}]{Nixon13}
{Nixon} C., {King} A., 2013, \apjl, 765, L7

\bibitem[{{O'Dea}(1985)}]{Odea85}
{O'Dea} C.~P., 1985, \apj, 295, 80

\bibitem[{{O'Dea} {et~al}\mbox{.}(2013){O'Dea}, {Baum}, {Tremblay}, {Kharb},
  {Cotton}, \& {Perley}}]{ODea13}
{O'Dea} C.~P., {Baum} S.~A., {Tremblay} G.~R., {Kharb} P., {Cotton} W.,
  {Perley} R., 2013, \apj, 771, 38

\bibitem[{{Orienti} \& {Prieto}(2010)}]{Orienti10}
{Orienti} M., {Prieto} M.~A., 2010, \mnras, 401, 2599

\bibitem[{{Poglitsch} {et~al}\mbox{.}(2010){Poglitsch}, {Waelkens}, {Geis}, \&
  {80 coauthors}}]{Poglitsch10}
{Poglitsch} A., {Waelkens} C., {Geis} N., {80 coauthors}, 2010, \aap, 518, L2

\bibitem[{{Rees}(1978)}]{Rees78}
{Rees} M.~J., 1978, \nat, 275, 516

\bibitem[{{Rush}, {Malkan} \& {Spinoglio}(1993){Rush}, {Malkan}, \&
  {Spinoglio}}]{Rush93}
{Rush} B., {Malkan} M.~A., {Spinoglio} L., 1993, \apjs, 89, 1

\bibitem[{{Schurch}, {Griffiths} \& {Warwick}(2006){Schurch}, {Griffiths}, \&
  {Warwick}}]{Schurch06}
{Schurch} N.~J., {Griffiths} R.~E., {Warwick} R.~S., 2006, \mnras, 371, 211

\bibitem[{{Shulevski} {et~al}\mbox{.}(2012){Shulevski}, {Morganti},
  {Oosterloo}, \& {Struve}}]{Shulevski12}
{Shulevski} A., {Morganti} R., {Oosterloo} T., {Struve} C., 2012, \aap, 545,
  A91

\bibitem[{{Sikora}, {Stawarz} \& {Lasota}(2007){Sikora}, {Stawarz}, \&
  {Lasota}}]{Sikora07}
{Sikora} M., {Stawarz} {\L}., {Lasota} J.-P., 2007, \apj, 658, 815

\bibitem[{{Stawarz}, {Sikora} \& {Ostrowski}(2003){Stawarz}, {Sikora}, \&
  {Ostrowski}}]{Stawarz03}
{Stawarz} {\L}., {Sikora} M., {Ostrowski} M., 2003, \apj, 597, 186

\bibitem[{{Terashima} \& {Wilson}(2003)}]{Terashima03}
{Terashima} Y., {Wilson} A.~S., 2003, \apj, 583, 145

\bibitem[{{Ulvestad}(2003)}]{Ulvestad03}
{Ulvestad} J.~S., 2003, in Astronomical Society of the Pacific Conference
  Series, Vol. 300, Radio Astronomy at the Fringe, {Zensus} J.~A., {Cohen}
  M.~H., {Ros} E., eds., p.~97

\bibitem[{{Ulvestad}, {Antonucci} \& {Barvainis}(2005){Ulvestad}, {Antonucci},
  \& {Barvainis}}]{Ulvestad05}
{Ulvestad} J.~S., {Antonucci} R.~R.~J., {Barvainis} R., 2005, \apj, 621, 123

\bibitem[{{Ulvestad} \& {Ho}(2001)}]{Ulvestad01}
{Ulvestad} J.~S., {Ho} L.~C., 2001, \apjl, 562, L133

\bibitem[{{Urry} \& {Padovani}(1995)}]{UrryPadovani95}
{Urry} C.~M., {Padovani} P., 1995, \pasp, 107, 803

\bibitem[{{van der Laan} \& {Perola}(1969)}]{vanderlaan69}
{van der Laan} H., {Perola} G.~C., 1969, \aap, 3, 468

\bibitem[{{Verdoes Kleijn} {et~al}\mbox{.}(1999){Verdoes Kleijn}, {Baum}, {de
  Zeeuw}, \& {O'Dea}}]{Verdoes99}
{Verdoes Kleijn} G.~A., {Baum} S.~A., {de Zeeuw} P.~T., {O'Dea} C.~P., 1999,
  \aj, 118, 2592

\bibitem[{{Verdoes Kleijn} {et~al}\mbox{.}(2002){Verdoes Kleijn}, {Baum}, {de
  Zeeuw}, \& {O'Dea}}]{Verdoes02}
{Verdoes Kleijn} G.~A., {Baum} S.~A., {de Zeeuw} P.~T., {O'Dea} C.~P., 2002,
  \aj, 123, 1334

\bibitem[{{Volonteri}, {Sikora} \& {Lasota}(2007){Volonteri}, {Sikora}, \&
  {Lasota}}]{Volonteri07}
{Volonteri} M., {Sikora} M., {Lasota} J.-P., 2007, \apj, 667, 704

\bibitem[{{Wandel}, {Peterson} \& {Malkan}(1999){Wandel}, {Peterson}, \&
  {Malkan}}]{Wandel99}
{Wandel} A., {Peterson} B.~M., {Malkan} M.~A., 1999, \apj, 526, 579

\bibitem[{{Wilson} \& {Ulvestad}(1982)}]{Wilson82}
{Wilson} A.~S., {Ulvestad} J.~S., 1982, \apj, 263, 576

\bibitem[{{Woo} \& {Urry}(2002)}]{Woo02}
{Woo} J.-H., {Urry} C.~M., 2002, \apj, 579, 530

\end{thebibliography}

\clearpage

\clearpage

\clearpage

\clearpage

\clearpage

\clearpage

\clearpage

\clearpage

\appendix
\section{Radio Loudness of a Selected Sample of FRI and Seyfert galaxies}
We present the radio loudness parameters of a set of FRI radio galaxies belonging to the UGC sample \citep{Verdoes99} and Seyfert galaxies belonging to the extended 12$\mu$m sample \citep{Rush93}, having taken into account their nuclear optical luminosities, in Table~\ref{tabradl}. We obtained the nuclear optical luminosities of the FRI radio galaxies from the F555W and F547W {\it HST} observations of \citet{Verdoes02}. These were converted to $B$-band (4400\AA) flux densities using an optical spectral index of $-0.9$ \citep[e.g.,][]{Chiaberge99}. A few FRI radio galaxies did not have {\it HST} data: total $B$-band magnitudes were used in their $R$ estimation instead. These have an asterisk next to their $R$ values. As most of the Seyfert galaxies considered here did not have {\it HST} data, we adopted the median nuclear absolute $B$-band magnitudes derived by \citet{HoPeng01}. Two sets of radio loudness parameters were derived using the median nuclear  absolute magnitudes for the CfA ($M_B^{nuc}=-17.4$) and Palomar ($M_B^{nuc}=-14.6$) Seyfert samples, respectively.
After deriving the apparent nuclear magnitudes ($m_B^{nuc}$) for the sources at different redshifts, the optical nuclear flux density at 4400\AA~($S_{o}$) was estimated using the relation: $S_{o}=10^{((-{m_B^{nuc}} - 48.36)/2.5)}$. The apparent $B$-band magnitudes and 5~GHz (=6cm) flux densities listed in Table~\ref{tabradl} were obtained from NED. We note that with the exception of a few FRI radio galaxies (the ones with an asterisk next to their $R$ values), these $B$-band magnitudes were not used in the actual estimation of $R$ for most sources, and are presented here for completeness.

We find that 66~percent of Seyfert 1s and 76~percent of Seyfert 2s are radio-loud taking into account the median nuclear magnitude of the CfA sample, while 89~percent of Seyfert 1s and 94~percent of Seyfert 2s are radio-loud taking into account the median nuclear magnitude of the Palomar sample. 17 Seyfert galaxies are common between the sample considered here and that for which nuclear $R$ values were estimated by \citet{HoPeng01}. The $R$ values of nearly 76~percent of the common sources (i.e., 13 of 17 sources) lie in the range that we have derived in Table~A1. The lower range of $R$ values derived here are 50 and 100 times higher than those obtained by Ho \& Peng for IZW1 and MK231, respectively, while the $R$ values derived by Ho \& Peng are 130 and 220 times higher than the upper range of $R$ values derived here for N2639 and N3031, respectively. This gives us confidence that the range of $R$ values derived for Seyfert galaxies here are truly representative of the sample.

\begin{table}
\caption{Radio Loudness Parameters}
\begin{tabular}{llllcc}
\hline\hline
{Source}&{B-band}&{Flux@6cm}&{$R$}&{Type}\\
{name}&{mag}&{mJy}&{}&{}\\
\hline
U00408       & 13.87  &  864.0    & 263788  & FRI  \\
U00597       & 12.23  &  1192.0   & 68375   & FRI  \\
U00689       & 13.38  &  2080.0   & 164054  & FRI  \\
U01004       & 13.49  &  77.6     & 19586   & FRI  \\
U01413       & 12.35  &  305.0    & 151932  & FRI  \\
U01841       & 13.75  &  2942.0   & 107096  & FRI  \\
U03695       & 13.60  &  339.0    & 3775    & FRI  \\
U05073       & 14.05  &  138.0    & 16326   & FRI  \\
U06635       & 12.57  &  635.0    & 15$^\ast$ & FRI  \\
U06723       & 13.52  &  1674.0   & 16678   & FRI  \\
U07115       & 14.33  &  169.0    & 9409    & FRI  \\
U07360       & 11.37  &  4089.0   & 2.0E+06 & FRI  \\
U07455       & 13.40  &  82.0     & 6209    & FRI  \\
U07494       & 10.29  &  3582.0   & 109362  & FRI  \\
U07654       & 9.65   &  59740.0  & 176693  & FRI  \\
U08419       & 12.90  &  449.0    & 14$^\ast$& FRI  \\
U08433       & 13.79  &  421.0    & 83886   & FRI  \\
U09058       & 12.80  &  325.0    & 918210  & FRI  \\
U11718       & 13.40  &  122.0    & 6$^\ast$ & FRI  \\
U12064       & 14.49  &  1146.0   & 158$^\ast$& FRI  \\
U12531       & 12.41  &  152.0    & 77764   & FRI  \\  
MK335        & 14.13  &  3.6      & 11, 141 & SY1  \\
IZW1         & 14.59  &  3.1      & 51, 670 & SY1   \\
M-3-7-11     & 13.94  &  31.5     & 163, 2149 & SY1   \\
N931         & 13.90  &  1.7      & 2, 28 & SY1   \\
N1365        & 10.23  &  230.0    & 30, 393 & SY1   \\
N1566        & 10.13  &  100.0    & 11, 144 & SY1   \\
MK618        & 16.59  &  5.0      & 29, 380 & SY1   \\
MK6          & 14.80  &  115.0    & 181, 2382 & SY1  \\
MK9          & 15.20  &  1.8      & 13, 173 & SY1  \\
MK79         & 13.30  &  10.0     & 22, 290 & SY1  \\
N2639        & 12.40  &  119.2    & 65, 855 & SY1  \\
MK704        & 15.20  &  4.0      & 15, 204 & SY1  \\
U5101        & 15.50  &  79.0     & 562, 7405 & SY1  \\
N2992        & 12.63  &  102.0    & 27, 350 & SY1  \\
MK1239       & 14.50  &  32.0     & 56, 744 & SY1  \\
N3031        & 8.10   &  385.0    & 0.02, 0.28 & SY1  \\
N3227        & 12.20  &  35.0     & 2, 30 & SY1  \\
N3516        & 12.30  &  10.6     & 4, 48 & SY1  \\
N4051        & 11.50  &  32.0     & 1, 10 & SY1  \\
N4151        & 11.20  &  152.0    & 7, 96 & SY1  \\
MK766        & 13.70  &  15.8     & 12, 153 & SY1  \\
N4579        & 11.50  &  99.0     & 11, 145 & SY1  \\
N4593        & 11.67  &  3.8      & 1, 18 & SY1  \\
\hline
\end{tabular}\\
{Column~1 lists the names of FRI radio galaxies from the UGC sample \citep{Verdoes99} and Seyfert galaxies from the Extended 12$\mu$m sample \citep{Rush93}, that were used for the current study. Columns~2 and 3 list their apparent $B$-band magnitudes and 5~GHz (=6cm) flux densities, respectively. These were obtained from NED. The nuclear optical luminosities for the FRI radio galaxies were obtained from the {\it HST} data of \citet{Verdoes02}. Column~4 lists the radio-loudness parameters, $R$. $\ast$ denotes those FRIs for which {\it HST} data were unavailable $-$ for these the B-band magnitudes listed {here} were used instead. As most Seyferts did not have {\it HST} data, two median nuclear absolute magnitudes of the CfA (=$-$17.4) and Palomar Seyfert samples \citep[=$-$14.6;][]{HoPeng01} were chosen, to derive two sets of $R$ values. {(Note: The $B$-band magnitudes listed here were not used in the estimation of $R$ for the Seyfert and most of the FRI galaxies.)}}
\label{tabradl}
\end{table}

\begin{table}
\contcaption{}
\begin{tabular}{llllcc}
\hline\hline
{Source}&{B-band}&{Flux@6cm}&{$R$}&{Type}\\
{name}&{mag}&{mJy}&{}&{}\\
\hline
N4594        & 8.56   &  156.0    & 8, 110 & SY1  \\
N4602        & 12.31  &  1.3      & 0.4, 5 & SY1  \\
M-2-33-34    & 17.50  &  6.0      & 5, 66 & SY1  \\
MK231        & 14.10  &  419.0    & 3433, 45258 & SY1  \\
N5033        & 10.90  &  82.0     & 3, 40 & SY1  \\
M-6-30-15    & 13.86  &  1.0      & 0.3, 4 & SY1  \\
N5548        & 13.10  &  13.8     & 18, 238 & SY1  \\
MK817        & 14.30  &  6.0      & 27, 355 & SY1  \\
MK509        & 11.83  &  5.5      & 29, 388 & SY1  \\
N7213        & 11.49  &  249.0    & 37, 488 & SY1  \\
N7469        & 13.00  &  95.0     & 112, 1478 & SY1  \\
N7603        & 14.40  &  20.0     & 79, 1039 & SY1  \\
N7674        & 13.52  &  86.0     & 2141, 28223 & SY2  \\
N7582        & 11.26  &  143.0    & 61, 807     & SY2  \\
I5063        & 12.86  &  524.0    & 4757, 62709 & SY2  \\
N6810        & 12.36  &  72.0     & 14, 190     & SY2  \\
U9913        & 13.90  &  208.0    & 746, 9829  & SY2  \\
N5953        & 13.30  &  40.0     & 26, 346  & SY2  \\
N5929        & 12.99  &  42.0     & 35, 463 & SY2  \\
N5506        & 13.60  &  227.2    & 92, 1219 & SY2  \\
MK463        & 14.80  &  118.0    & 14473, 190789 & SY2  \\
N5347        & 13.30  &  3.1      & 3, 36 & SY2  \\
MK273        & 15.00  &  103.0    & 1677, 22107 & SY2  \\
N5256        & 14.10  &  49.0     & 818, 10781 & SY2  \\
N5194        & 8.80   &  1135.0   & 22, 295 & SY2  \\
N5135        & 12.76  &  107.0    & 291, 3835 & SY2  \\
N5005        & 10.60  &  72.0     & 22, 284 & SY2  \\
N4941        & 12.10  &  9.0      & 0.5, 7 & SY2  \\
N4501        & 10.60  &  102.4    & 189, 2498 & SY2  \\
N4388        & 12.20  &    93.0   & 54, 709 & SY2  \\
N3982        & 11.60  &    24.0   & 8, 104 & SY2  \\
N3660        & 12.69  &   0.8     & 0.5, 7& SY2   \\
N3079        & 11.20  &    327.0  & 122, 1604 & SY2  \\
N1667        & 13.12  &   45.0    & 135, 1780 & SY2  \\
N1386        & 12.24  &    13.0   & 3, 45 & SY2  \\
N1320        & 13.32  &    3.3    & 5, 61 & SY2  \\
N1241        & 12.64  &    12.3   & 10, 129 & SY2  \\
N1194        & 14.70  &    1.5    & 3, 45 & SY2 \\                         
N1097        & 9.96   &    150.0  & 90, 1182 & SY2  \\
N1068        & 9.70   &    2187.0 & 672, 8858 & SY2  \\
N1056        & 13.50  &    4.8    & 34, 442 & SY2  \\
F01475-0740  & 17.13  &    278.0  & 502, 6624 & SY2  \\
N513         & 13.40  &    57.0   & 83, 1100 & SY2  \\
N262         & 15.00  &    681.0  & 125, 1649 & SY2  \\
\hline
\end{tabular}
\end{table}
\end{document}